\documentclass{emulateapj}
\usepackage{graphicx}

\newcommand{\hi}          {\mbox{\rm \ion{H}{1}}}
\newcommand{\hii}         {\mbox{\rm \ion{H}{2}}}

\newcommand{\um}          {$\mu$m}
\newcommand{\ha}          {H$\alpha$}
\newcommand{\hh}          {\hbox{$^{\rm h}$}}
\newcommand{\hm}          {\hbox{$^{\rm m}$}}

\newcommand{\ftf}         {\mbox{F$_{24}$}}
\newcommand{\fei}          {\mbox{F$_{8}$}}
\newcommand{\fst}          {\mbox{F$_{70}$}}

\begin{document}

\title{The \emph{Spitzer} Survey of the Small Magellanic Cloud:
  S$^3$MC Imaging and Photometry in the Mid- and Far-Infrared Wavebands} 

\author{Alberto D. Bolatto\altaffilmark{1}, Joshua
D. Simon\altaffilmark{2}, Sne\v{z}ana
Stanimirovi{\'c}\altaffilmark{1}, Jacco Th. van Loon\altaffilmark{3},
Ronak Y. Shah\altaffilmark{4}, Kim Venn\altaffilmark{5}, Adam
K. Leroy\altaffilmark{1}, Karin Sandstrom\altaffilmark{1}, James
M. Jackson\altaffilmark{4}, Frank P. Israel\altaffilmark{6}, Aigen
Li\altaffilmark{7}, Lister Staveley-Smith\altaffilmark{8}, Caroline
Bot\altaffilmark{2}, Francois Boulanger\altaffilmark{9}, \& M\'onica
Rubio\altaffilmark{10}}

\altaffiltext{1}{Department of Astronomy and Radio Astronomy
Laboratory, University of California at Berkeley, 601 Campbell Hall,
Berkeley, CA 94720, USA}

\email{bolatto@berkeley.edu} 

\altaffiltext{2}{Department of Astronomy, California Institute of
Technology, Pasadena, CA 91125, USA}

\altaffiltext{3}{Astrophysics Group, Lennard Jones Laboratories,
Keele University, Staffordshire ST5 5BG, UK}

\altaffiltext{4}{Department of Astronomy and Center for Astrophysical
Research, Boston University, Boston, MA 02215, USA}

\altaffiltext{5}{Department of Astronomy, University of Victoria,
Victoria BC V8W 2Y2, Canada}

\altaffiltext{6}{Leiden Observatory, NL-2300 RA Leiden, The Netherlands}

\altaffiltext{7}{Department of Physics and Astronomy, University of
Missouri-Columbia, Columbia, MO 65211, USA}

\altaffiltext{8}{Australia Telescope National Facility, Epping NSW
1710, Australia}

\altaffiltext{9}{Institut d'Astrophysique Spatiale, Universit\'e
Paris-Sud 11, 91405 Orsay, France}

\altaffiltext{10}{Departamento de Astronom\'{\i}a, Universidad de
Chile, Casilla 36-D Santiago, Chile}

\begin{abstract}
We present the initial results from the {\em Spitzer} Survey of the
Small Magellanic Cloud (S$^3$MC), which imaged the star-forming body
of the Small Magellanic Cloud (SMC) in all seven MIPS and IRAC
wavebands.  We find that the \fei/\ftf\ ratio (an estimate of PAH
abundance) has large spatial variations and takes a wide range of
values that are unrelated to metallicity but anticorrelated with 24
\um\ brightness and $\ftf/\fst$ ratio.  This suggests that
photodestruction is primarily responsible for the low abundance of
PAHs observed in star-forming low-metallicity galaxies. We use the
S$^3$MC images to compile a photometric catalog of $\sim400,000$ mid-
and far-infrared point sources in the SMC.  The sources detected at
the longest wavelengths fall into four main categories: 1) bright
$5.8$ \um\ sources with very faint optical counterparts and very red
mid-infrared colors ($[5.8]-[8.0]>1.2$), which we identify as YSOs. 2)
Bright mid-infrared sources with mildly red colors
($0.16\lesssim[5.8]-[8.0]<0.6$), identified as carbon stars. 3) Bright
mid-infrared sources with neutral colors and bright optical
counterparts, corresponding to oxygen-rich evolved stars. And, 4)
unreddened early B stars (B3 to O9) with a large 24 \um\ excess. This
excess is reminiscent of debris disks, and is detected in only a small
fraction of these stars ($\lesssim5\%$). The majority of the brightest
infrared point sources in the SMC fall into groups one to three. We
use this photometric information to produce a catalog of 282 bright
YSOs in the SMC with a very low level of contamination ($\sim7\%$).
\end{abstract}


\section{Introduction}

The Small Magellanic Cloud (SMC), one of the closest and most
prominent neighbors of the Milky Way, is a southern hemisphere dwarf
galaxy of low mass \citep[M$_{dyn}\sim2.4\times10^9$
M$_\odot$;][]{stanimirovic04} and small size (R$_*\sim3$ kpc).  The
SMC is actively interacting with its companion the Large Magellanic
Cloud \citep[LMC, studied by the {\em Spitzer} SAGE
project;][]{meixner06}. Both galaxies are being tidally disrupted by
the Milky Way, and also suffer from ram-pressure interactions with the
halo of hot gas surrounding the Galaxy \citep[e.g.,][]{gardiner96}.
The results of this complex dynamical interplay are the dramatic
features of the Magellanic Bridge and Magellanic Stream, which
dominate the high-velocity \hi\ emission in the Southern sky.

Befitting its nature as a gas-rich late-type dwarf galaxy, the SMC
contains a number of sites of active star formation that are apparent
in \ha\ images, although none are as prominent as the 30 Doradus
starburst in the LMC.  The brightest SMC star forming complex is the
giant \hii\ region N~66 \citep[also known as DEM~103, or
NGC~346;][]{henize56,dem76}, which contains approximately $60$ O stars
\citep{mpg89}. The star formation activity of the SMC, as well as most
of its gas, are concentrated in the feature known as the ``Bar''
(probably the body of the SMC proper). A second feature, the ``Wing'',
is an extension roughly perpendicular to the Bar on its eastern side
that continues into the \hi\ bridge that joins both Clouds
(Figure \ref{coverage}).  As one may expect in an interacting system, the
three-dimensional structure of the galaxy remains quite unclear.
Recent analyses show that the prominent irregular features, which are
apparent in the light of young stars, disappear when looking at the
old stellar populations
\citep{cioni00,zaritsky00,maragoudaki01}. These are regularly
distributed, and the shape of the SMC appears spheroidal. Observations
indicate that the distances to the Bar and the Wing are measurably
different, and that the line-of-sight depth of the SMC is not
negligible, amounting to 6 to 12 kpc \citep[][and references
therein]{crowl01}. The distance to the center of the SMC is
approximately 61.1 kpc, corresponding to a distance modulus of
$m-M=18.93$~mag and a spatial scale of 0.3 pc/\arcsec
\citep{hilditch05,keller&wood06}. To put this in context, the spatial
resolution of the $3.6-8.0$ \um\ data presented here is similar to
the 60 \um\ resolution of the {\em Infrared Astronomical Satellite}
(IRAS) at 2 kpc.  The thickness of the SMC introduces a dispersion of
$\sim0.2$~mag in the distance modulus for objects located near either
edge. The line-of-sight to the SMC is very clear, with a foreground
extinction $A_V\sim0.12$, corresponding to a reddening of
$E(B-V)\sim0.04$ \citep{schlegel98}.

The proximity of the SMC makes it the lowest metallicity gas-rich
galaxy that can be studied in great detail. For observational reasons
metallicity, $Z$, is commonly quantified using the oxygen abundance
relative to hydrogen. The gas-phase oxygen abundances measured in SMC
\hii\ regions are $12+\log{\rm(O/H)}\approx8.0$
\citep{dufour75,dufour77,dufour84}, only a factor of 4 to 5 higher
than those observed in the lowest metallicity galaxies known (I~Zw~18
and SBS~0335-052), and a factor of 5 lower than solar metallicity.
The same metallicity with very little dispersion is observed in
several stellar types in the SMC, including K, F, and A supergiants
and Cepheid variables \citep{russell89,hill97,luck98,venn99}. Its low
heavy-element abundance suggests that the SMC may be one of the best
local templates for studying the primitive interstellar medium and
providing insight into the processes at work in primordial
galaxies at high redshift. In particular, the interaction between
massive stars, gas, and dust is unusually accessible in this nearby
galaxy.

Compared to the Milky Way, the dust properties of the SMC are
peculiar. For quite a while it has been recognized that the UV
extinction curve in the SMC lacks the 2175 \AA ``hump''
\citep[e.g.,][]{gordon03} which is interpreted as signaling the lack
of a population of small carbonaceous dust grains \citep{li02}.
Indeed, most current models of dust composition in the SMC rely
heavily on silicates to reproduce its extinction curve, with typical
silicate to carbon ratios by mass in the range $2-12$
\citep[e.g.,][]{weingartner01}.  Whether this dearth of carbonaceous
grains is maintained for the large grains remains unclear, as there is
a well known anticorrelation between the specific frequency of carbon
stars (the main producers of carbonaceous grains) and galactic
metallicity \citep[e.g.,][]{blanco80}, that results in a large
abundance of these stars in the SMC.

Other lines of research also suggest a dearth of the smallest dust
grains in the low-metallicity interstellar medium of the SMC. The
first wide-field infrared maps of the SMC were acquired by IRAS and
presented by \citet{si89}.  \citet{sauvage90} studied the IRAS colors
of the SMC, comparing them to the emission from larger, higher
metallicity galaxies, and concluded that there was evidence for a
paucity of the small grains that are responsible for the bulk of the
extended 12 $\mu$m emission. More recently, \cite{snez00} reprocessed
the IRAS data to produce maps with improved angular resolution of
1\arcmin\ at 12~\um, 25~\um, and 60~\um, and 1\farcm7 at 100~\um.  By
comparing these data with \hi\ observations of the SMC at similar
resolution, they were able to study the large-scale properties of the
ISM.  \citeauthor{snez00} used these far-infrared (FIR) observations
to measure the dust mass, the dust temperature and its spatial
variations, and the dust-to-gas ratio, DGR, across the SMC.  This
study found a mean dust-to-gas ratio in the SMC that is a factor of
$\sim100$ times lower than in the Milky Way, providing a different
environment for star formation to take place.  It also concluded that
the dust in the SMC is dominated by large grains, with weaker
contributions from polycyclic aromatic hydrocarbons \citep[PAHs,
typical sizes 4\AA$<a<$12\AA;][]{desert90} and very small grains
(VSGs, 12\AA$<a<$150\AA) than are found in the Galaxy. One of the
limitations of this study is the use of the 60 \um\ IRAS waveband to
determine dust temperature. This waveband is expected to have a
substantial contribution from stochastically heated small dust grains,
out of equilibrium with the radiation field.

\citet{bot04} expanded this work by adding {\emph{Infrared Space
Observatory}} (ISO) 170~\um\ imaging of the SMC to the IRAS data.
Using a combination of 100 and 170 \um\ data they measured an
appreciably higher dust-to-gas ratio, $\sim30$ times lower than
Galactic. They also modeled the spectral energy distribution to
decompose the emission into the contributions by PAHs, VSGs, and large
grains, finding a substantial 60 \um\ excess. Such excess could be
caused by an enhanced interstellar radiation field in the SMC, or
perhaps by a change in the grain size distribution with respect to the
Galaxy, with more VSGs relative to large grains. \citeauthor{bot04}
suggest that grain destruction, perhaps shattering by supernovae
shocks, is the driver behind the different properties of the dust in
the SMC and the Milky Way.

These studies point to a dust-to-gas ratio in the SMC much lower than
Galactic, scaling with its metallicity as DGR~$\propto Z^{2}$, a result
similar to that found for a collection of galaxies by
\citet{lisenfeld98}. Note that measurements of the mass-loss rate of
the most evolved asymptotic giant branch (AGB) stars in the SMC,
however, find a much larger dust-to-gas ratio, scaling as DGR~$\propto
Z^{1}$ \citep{vanloon06}. Because these stars are believed to dominate
the ISM dust production, this observation suggests that efficient
destruction of dust takes place in the metal-poor star-forming ISM,
along the lines of the evidence mentioned in the previous
paragraphs. 

Detailed modeling of SMC UV absorption and FIR emission data by
\citet{weingartner01}, \citet{li02}, and \citet{clayton03} indicate
that a deficit of PAHs, as well as the lower abundance of VSGs
suggested by IRAS, and perhaps a change in composition with more
silicate and fewer graphite grains are required to reproduce the
observations. \citet{contursi00} studied N~66 using the cameras
onboard ISO, and found that the emission from the Aromatic Infrared
Band (AIB) in the 6.75 $\mu$m ISO waveband, commonly attributed to
PAHs, was faint compared to the 15 $\mu$m dust continuum. This was
confirmed spectroscopically using ISO's continuously variable filter
(CVF) to obtain low-resolution spectra toward emission peaks, all of
which show very weak or nonexistent AIB emission. As a counterpoint,
\citet{reach00} used the CVF to obtain spectroscopy toward a quiescent
molecular cloud in the SMC. They measured AIB-to-continuum ratios
similar to those observed in the Milky Way, pointing to similar
relative abundances of PAHs.  These data suggest that PAHs do form and
survive in the SMC and other star-forming low-metallicity galaxies in
quantities very similar to Milky Way PAHs, where the environment
allows it. Indeed, SMC observations show us that dust properties are
environment dependent, and one of the main motivations for a new
infrared survey of the SMC is to study and characterize the influence
of the local environment on the properties and composition of
interstellar dust.

The \emph{Spitzer Space Telescope} allows dramatic improvements in our
ability to study the ISM of the SMC, with previously unattainable
images at 3.6 to 8.0 \um\ and a factor of $4-10$ higher angular
resolution and typically an order of magnitude better sensitivity at
FIR wavelengths than previous missions.  The {\em Spitzer} data enable
us to take a giant step forward in the study of the detailed,
distribution of all types of dust grains, including PAHs which
primarily emit at wavelengths shorter than 12~\um.  The high angular
resolution of \emph{Spitzer} means that point sources can be separated
from diffuse emission out to 70~\um, allowing investigations of large
samples of IR point sources (such as carbon stars and young stellar
objects) and simultaneously providing a cleaner picture of the ISM.
The unprecedented sensitivity and resolution of the {\em Spitzer} data
makes it possible to determine the abundance and distribution of dust
grains and PAHs in relationship to their environment, to test and
constrain dust models, and to obtain a complete census of star
formation in the SMC, the nearest ``primitive'' galaxy.

In this paper we present an overview of the \emph{Spitzer} Survey of
the Small Magellanic Cloud (S$^3$MC), discussing the observations,
methodology, and some of the initial results on the extended emission
and the infrared point sources in the SMC. This is the second in a
series of papers presenting the results of the survey \citep[see
also][]{snez05}. In \S\ref{obs} we describe the observations, in
\S\ref{datared} we describe in detail the data reduction process
employed to obtain the final S$^3$MC images, in \S\ref{photometry} we
discuss the point source photometry for the global survey and the
comparison with existing datasets, in \S\ref{discussion} we analyze
the extended dust emission and discuss the different stellar
populations detected by {\em Spitzer}, and in \S\ref{conclusion} we
summarize our findings.

\section{Observations}
\label{obs}

\begin{figure*}
\plotone{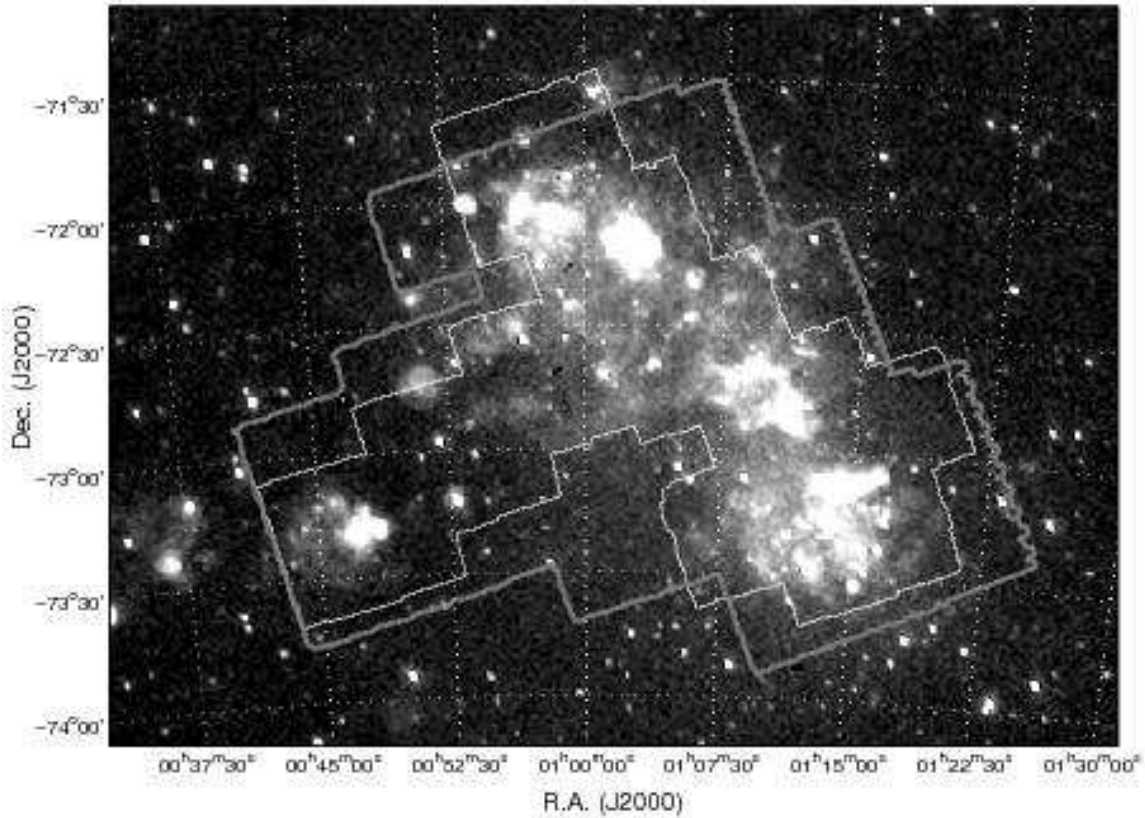} 
\figcaption{Spatial coverage of the S$^3$MC data, overlaid on the
  H$\alpha$ ``parking lot camera'' image of the SMC
  \protect\citep{kennicutt95}. The white outline indicates the
  coverage of the IRAC observations, while the gray outline shows the
  coverage for MIPS. Note that this is only approximate, since the
  placement of the cameras on the {\em Spitzer} focal plane results in
  slightly different areas of the sky mapped for the different
  wavebands. The SMC Bar is the region where the bulk of the star
  formation (indicated by H$\alpha$ emission) takes place, extending
  roughly from ($01\hh15\hm$,$-73^\circ30\arcmin$) to
  ($00\hh54\hm$,$-72^\circ00\arcmin$).  The Wing region, notable in \hi\
  but with very little diffuse H$\alpha$ emission, is approximately
  perpendicular to the bar and extends eastward from
  ($01\hh00\hm$,$-72^\circ45\arcmin$) to
  ($00\hh45\hm$,$-73^\circ15\arcmin$) and beyond. \label{coverage}}
\end{figure*}

We obtained {\it Spitzer Space Telescope} images of the Small
Magellanic Cloud in all the bands of the Infrared Array Camera
\citep[IRAC,][]{fazio04} and the Multiband Imaging Photometer and
Spectrometer \citep[MIPS,][]{rieke04} as part of the Cycle I General
Observer project 3316, ``The Small Magellanic Cloud: A Template for
the Primitive Interstellar Medium''.  This project was designed to map
as large an area of the SMC as possible with good sensitivity and
within the scope of a small \emph{Spitzer} proposal (less than 50
hours), with the goal of sampling a diverse set of environments. The
images cover an area of approximately 2.8 to 4 deg$^{2}$ in the
various bands, with a region of $\sim2.5^\circ\times1^\circ$ mapped along
the SMC Bar, and $\sim1.5^\circ\times1^\circ$ along the Wing, and the
Astronomical Observing Requests (AORs) were designed to provide
contiguous coverage without imposing scheduling constraints (Figure
\ref{coverage}).

The MIPS data were acquired in November 2004, using five large AORs
each lasting over $3$~hr, and two smaller AORs slightly over 1 hour
long each.  The angular resolution of {\em Spitzer} at 24, 70, and
160~$\mu$m is 6\arcsec, 18\arcsec, and 40\arcsec\ respectively. The
observations used the MIPS Scan Astronomical Observing Template (AOT),
with a medium ($6\farcs5$ s$^{-1}$) scan rate as a compromise between
coverage and sensitivity.  A total of over 11,700 individual images
were obtained in each of the three MIPS bands.  The 1-$\sigma$
sensitivities measured in the MIPS images are approximately 0.045 MJy
sr$^{-1}$ at 24~$\mu$m, 0.6 MJy sr$^{-1}$ at 70~$\mu$m, and 0.7 MJy
sr$^{-1}$ at 160 $\mu$m. The sensitivity is degraded at 70 $\mu$m by
the pattern noise present in the image.  At 160 $\mu$m it is
questionable whether it is possible to find regions of the image that
are free of emission in which to measure the noise level, so the
sensitivity estimate for this wavelength has a significant
uncertainty. Furthermore, to accommodate the observations within the
scope of the proposal the oversampling factor (i.e., the number of
times a given position in the sky is observed with different parts of
the detector) at 160 $\mu$m was two rather than the recommended four.
The corresponding 1-$\sigma$ point source sensitivities are 0.043, 5,
and 30 mJy respectively.

The IRAC data were obtained during 2005 May using 8 large 4 hour-long
and 2 small 2 hour-long AORs with a frame time of 12 seconds and a 3
point cycling dithering pattern.  The point response function of IRAC
has a FWHM of approximately $1\farcs66$, $1\farcs72$, $1\farcs88$, and
$1\farcs98$ in the 3.6, 4.5, 5.8, and 8.0 \um\ bands.  The typical
oversampling factor of a given position in the sky was
$\sim12$. Anticipating a number of bright point sources, the AOTs for
the fields along the SMC Bar were obtained using the high
dynamic range mode, which takes a short 0.6 second exposure along with
each 12 second exposure on the sky.  This allows us to recover the
correct fluxes for very bright stars that are saturated in the 12
second exposures. The saturation limits for stars in an 0.6 second
exposure are 630, 650, 4600, and 2500 mJy in the 3.6, 4.5, 5.8, and
8.0 $\mu$m bands respectively. Approximately 6,700 individual images
were obtained at each band.  The 1-$\sigma$ sensitivities estimated
from the mosaics are 0.008, 0.01, 0.04, and 0.04 MJy sr$^{-1}$,
corresponding to point source sensitivities of 0.7, 1.1, 8, and 6.5
$\mu$Jy respectively.  Table \ref{surveyprop} summarizes the survey
properties.

\begin{deluxetable*}{lcccccccc}
\tablecolumns{9}
\tablewidth{0pt}
\tablecaption{Survey Properties}
\tablehead{
\colhead{Property} & \colhead{Unit} & \colhead{3.6 \um} & \colhead{4.5 \um} & \colhead{5.8 \um} 
& \colhead{8.0 \um} & \colhead{24 \um} & \colhead{70 \um} & \colhead{160 \um}}
\startdata
Sensitivity\tablenotemark{1} & MJy sr$^{-1}$ & 0.008 & 0.010 & 0.040 & 0.040 & 0.045 & 0.600 & 0.700\\
Sensitivity\tablenotemark{2}& $\mu$Jy      & 0.7   & 1.1   & 8     & 6.5   & 43    & 5,000  & 30,000\\
Resolution  & arcsec       & 1.66  & 1.72  & 1.88  & 1.98  & 6     & 18    & 40 \\
Saturation  & mJy          & 630   & 650   & 4,600  & 2,500  & 1,000 & 5,750   & 750 \\
Area         & deg$^2$     & 2.77  & 2.77 & 2.77 & 2.77 & 3.94 & 3.70 & 3.69 \\
Source Count\tablenotemark{3}&     & 292,000 & 289,000 & 80,000 & 62,000 & 16,000 & 1,800 & \ldots \\
Completeness Flux\tablenotemark{4}& $\mu$Jy     & 31 & 24 & 80 & 80 & 400 & 40,000 & \ldots \\
Photometric Zero Point\tablenotemark{5}& Jy & 277.1 & 179.4 & 113.9 & 63.1 & 7.253 & 0.8198 & \dots \\
\enddata
\tablenotetext{1}{For extended emission, 1-$\sigma$.}
\tablenotetext{2}{For point sources, 1-$\sigma$.}
\tablenotetext{3}{Approximate.}
\tablenotetext{4}{Approximate. See discussion in \S\protect\ref{completeness}.}
\tablenotetext{5}{Martin Cohen, priv. comm.. Consistent with \protect\citet{reach05}.}
\label{surveyprop}
\end{deluxetable*}

\section{Data Reduction}
\label{datared}

We used the MOPEX data reduction package (version 050905) provided by
the {\it Spitzer} Science Center (SSC) to combine the individual Basic
Calibrated Data (BCD) frames into large mosaics.  There are a number
of possible methods for processing the data into a final mosaic of the
SMC.  We chose to begin with the BCD frames, which are derived from a
single data collection event (a DCE, or a single frame exposure).  The
BCD images were generated by the SSC using the standard pipeline
(version S11.4).  The detailed pipeline processing of raw images from
{\it Spitzer} to BCD FITS files is fully discussed by
\citet{gordon05} and \citet{masci2005}.  Each BCD image is
accompanied by an uncertainty and a mask image, which allow proper
weighting of the data and removal of corrupt pixels.

The SSC pipeline provides, in addition to the BCD images, a
preliminary mosaic of each AOR referred to as the post-BCD mosaic.  We
use these images primarily for comparison when we generate the full,
much larger mosaic of the SMC.

\subsection{MIPS Data Reduction}

Inspection of the 24$~\mu$m MIPS BCD and post-BCD images revealed a
small number of saturated sources and obvious latent images from
bright sources or cosmic ray hits.  No attempt was made to explicitly
remove cosmic rays from individual BCD images before combining frames
together. To avoid including cosmic ray hits in the final mosaics, we
used the redundancy and outlier rejection in mosaicking.  Only
temporal outlier rejection was performed, where redundant fields taken
at different times are compared to remove artifacts.

The 70~$\mu$m data contain several instrumental effects as well as
bright extended emission that make point-source extraction
difficult. The SSC provides two types of BCD images: normally
processed (unfiltered) and those with a time-median filter applied
that removes most of the background signal (filtered). Following the
recommendation of the SSC, we used the 70~$\mu$m filtered data to
obtain point-source fluxes. However, only the unfiltered data preserve
extended emission.  In this paper, we present mosaics made from the
unfiltered 70~$\mu$m data.  Because of the complexities of dealing
with the ``first frame'' effect, row-droop, and other similar
artifacts in the current software pipeline \citep{gordon05}, we have
not applied any extensive processing to form the 70~$\mu$m mosaic.


We combined the individual MIPS observations to form a mosaic using
the default linear interpolation method. Adjacent, partially
overlapping fields were combined using corrections computed by the
MOPEX task {\tt overlap.pl}, which matches fluxes in common regions to
guarantee a consistent background level. Because this process is
computer intensive, we applied it to subsets of $\sim$3,000 images at
a time. MOPEX calculates the offsets corresponding to each image 
by generating a matrix of median values for different DCEs and
solving a system of linear equations.

We linearly combined the background-corrected MIPS BCD images,
weighting them by the number of coverages at a given position and the
associated uncertainty image.  The outlier rejection settings employed
are described above.

\subsection{IRAC Data Reduction}

For the IRAC data, we performed a number of pre-processing steps on
the individual BCDs before combining them into a complete mosaic.  We
used the full ``dual-outlier'' rejection method (see following
paragraphs) in order to address bright, solar system objects whose
proper motion is easily detectable during our dwell time on the SMC.
Following recommendations from the SSC (D. Makovoz, priv. comm.)  we
used the default linear interpolation method to resample pixels from
the BCD images into the new fiducial mosaic image frame in all
instances.

Following this correction, we used specialized MOPEX routines to
remove two artifacts present in the IRAC images: the so called
muxbleed and column-pulldown cosmetic effects caused by the presence
of bright sources in individual DCEs.  Muxbleed manifests itself as a
trail of repeating bright pixels (every fourth pixel) along the
readout direction of the chip after detection of an extremely bright
star. The effect of column-pulldown is to bring down the counts in an
entire column associated with a bright source \citep{patten2004}.
These artifacts are dealt with using the MOPEX script {\tt
cosmetic.pl}.  Pixels are masked out of the final image processing if
both 1) at least one pixel is found to exceed 50 MJy/sr and 2) if
more than 10\% of the pixels along the bright pixel row are
found to have fluxes outside of the range 2--40$\times$ the RMS
value for the row.  These settings are the default values,
found to work best in removing the muxbleed problems.

The next pre-processing step removed the saturated pixels in those
fields (along the Bar) for which we obtained high dynamic-range
imaging. We compared the short and long exposures for each pixel and
identified saturated pixels. Those pixels found to be saturated in the
long exposure were then replaced with the corresponding pixel in the
short exposure. 


An important component of the mosaicking of the IRAC BCDs is outlier
rejection.  We combined the corrected IRAC BCD images by performing
both time-domain (outlier rejection) and the more elaborate
spatial-temporal (dual-outlier rejection) filtering processes.  We used
the recommended (default) settings for the values of the relevant
modules in outlier rejection. We linearly interpolated the BCD data
onto the final mosaic. The images are combined as a weighted sum,
using the associated uncertainty images, to form the final image.

The IRAC images from the 5.8 \um\ waveband (and, to a lesser extent,
those at 8.0 \um) presented an additional challenge in the form of
large-scale additive gradients in the background throughout the
mosaic. The gradient is caused by the combination of a time varying
offset in the readout amplifier, and the mosaicking algorithm
implemented in MOPEX (SSC, priv. comm.).  Following the
recommendations of the SSC, we made a ``super dark'' frame using the
BCDs within each AOR.  Using procedures developed in IDL, we stacked
the BCDs for each AOR and computed a pixel-by-pixel median value. We
then subtracted the overall median of this super dark from each pixel,
so that its DC level was zero. We subtracted the super dark from each
BCD, flat fielded the images, and continued with the same mosaicking
routines that were used for the 3.6 and 4.5 \um\ wavebands.

\subsection{{\em Spitzer} Flux Calibration and Color Corrections}

The dominant process that affects the IRAC calibration is the
location-dependent photometric flat-field corrections for compact
sources, caused by geometric array distortions and by the change in
the effective filter bandpass as a function of the incidence angles of
the radiation \citep{quijada2004}.  This effect can amount to a 10\%
peak-to-peak calibration error across the array.  The ultimate
absolute point source calibration attainable by IRAC is $\approx2\%$
\citep{reach05}. Because of the dithering and oversampling used to
obtain our images, we believe their relative calibration may be closer
to this latter limit. The complex backgrounds and the level of
confusion in dense stellar regions of the SMC, however, are probably
the limiting factors in our photometric accuracy.

The MIPS flux densities have an estimated absolute uncertainty of 10\%
and 20\%, for 24 and 70 \um, respectively \citep{rieke04}. These
uncertainties are dominated by the uncertainty in the fluxes of the
calibration sources. 

To convert our photometric fluxes to the Vega magnitud system we use
the zero points determined by M. Cohen (priv. comm.), which are
consistent with \citet{reach05}.  They are listed in Table
\ref{surveyprop}. 

The definition of monochromatic flux density used by IRAC, which is
common to IRAS and MSX, assumes an underlying flux density
$F_\nu\propto\nu^{-1}$. The MIPS flux scale, however, assumes a
Rayleigh-Jeans underlying flux density $F_\nu\propto\nu^2$. To convert
the MIPS scale into the IRAC scale it is necessary to apply factors of
1.041, 1.089, and 1.043 to the MIPS fluxes measured in the 24, 70, and
160 \um\ wavebands respectively. In the magnitude scale, these color
corrections correspond to -0.043, -0.094, and -0.046 mag
respectively. Note that {\em throughout this paper we use the flux
definitions native to each instrument}. Thus these corrections are not
applied.

\section{Point Source Photometry}
\label{photometry}

Point source photometry was carried out on the mosaic images using the
Astronomical Point Source Extractor (APEX) software in the MOPEX
package (version 050905) provided by the SSC.  Using the routine {\tt
apex\_1frame}, we detected bright sources in the images with a default
point response function (PRF) generated from in-orbit checkout
observations.  We selected $\sim30$ of the most isolated of these
stars, covering a range of brightnesses and positions, and built a new
PRF for each band with the {\tt prf\_response} routine.  We then
fitted the new PRF to the data and calculated flux densities for all
sources detected at a signal-to-noise ratio greater than 5 (again with
{\tt apex\_1frame}).  Aperture correction factors of 1.12, 1.01, 0.91,
0.92, 0.98, and 1.00 were applied to the 3.6, 4.5, 5.8, 8.0, 24, and
70 \um\ PRF fitting photometry respectively, to bring the measured flux
densities into agreement with very large aperture measurements of
chosen bright stars.  Finally, we created a point source-subtracted
image with {\tt apex\_qa} to assess the results of the PRF fitting and
to examine the extended emission that remained.

\begin{figure*}
\includegraphics[width=7.1in]{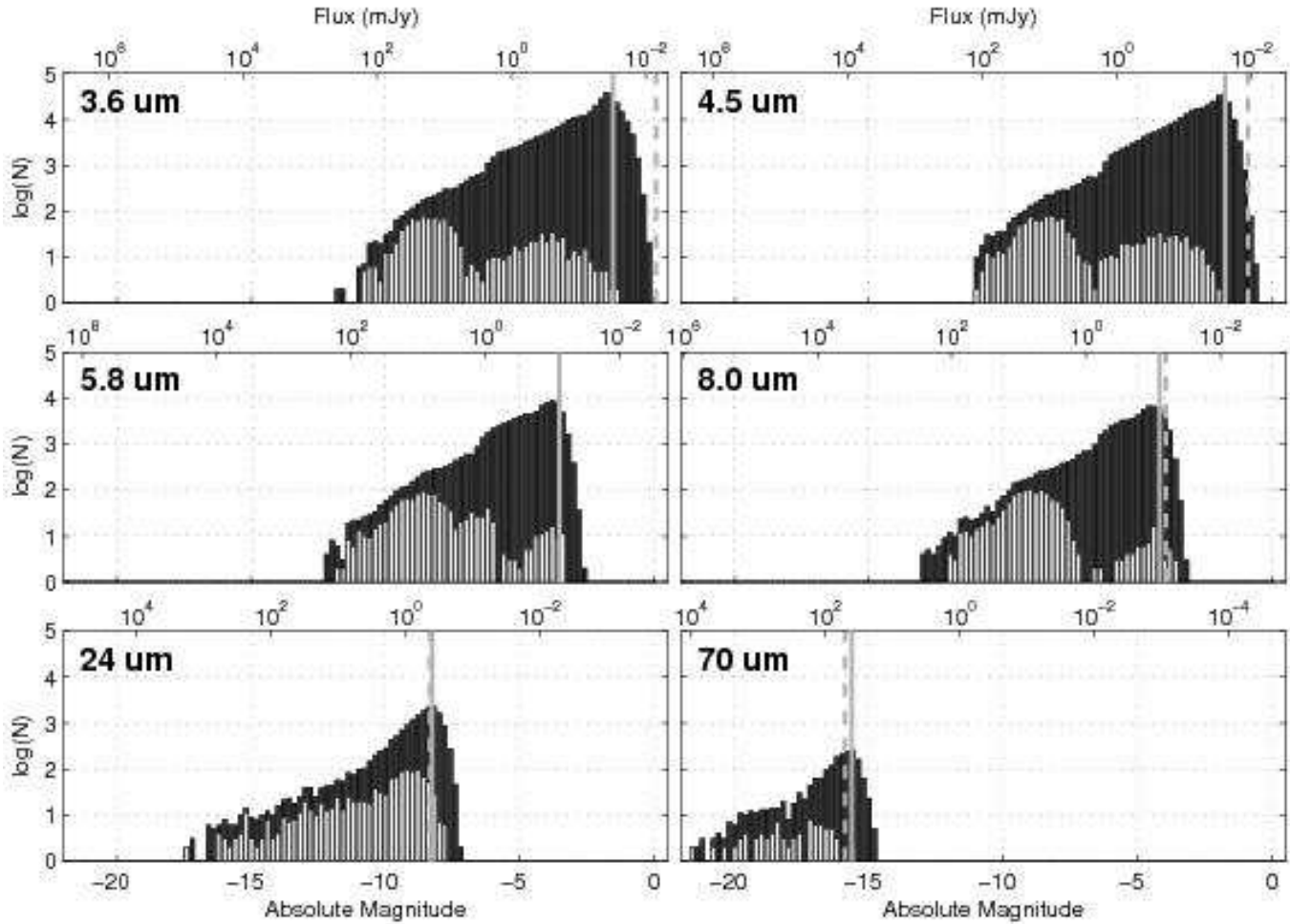}
\figcaption{Source count histograms for the different {\em Spitzer}
  wavebands in black (bin size is 0.2 mag). The bottom axis scale,
  stellar magnitudes, is shared between panels. The top axis scale,
  stellar flux, is different in each panel due to the different
  magnitude zero points.  The photometric completeness thresholds
  estimated based on 10 times the sensitivities of the images are
  indicated by the dashed vertical lines: 7, 11, 80, 65, 430, and
  50000 $\mu$Jy for the 3.6 to 70 \um\ wavebands respectively. The
  solid lines indicate the peak of the source count histograms, below
  which the photometry is clearly incomplete: 31, 24, 80, 80, 400, and
  40000 $\mu$Jy respectively.  The completeness of the photometry in
  the 3.6 and 4.5 \um\ wavebands, where the source densities are
  higher and confusion is more severe, is consequently worse than
  expected based on the sensitivity of the images alone. The white
  histograms show the source counts for the classes of sources
  identified in \S\protect\ref{starpops}. A large fraction
  ($\sim50\%$) of the brightest sources in each waveband fall into one
  of the identified classes.
\label{photohist}}
\end{figure*}

We can use the source count histograms to estimate the fluxes at which
incompleteness in the photometry becomes significant.  Figure
\ref{photohist} shows the histograms of the counts for a given source
flux in each {\em Spitzer} waveband. Assuming that the intrinsic
source counts approximately follow a power law, we use the peak of the
histogram to find the flux below which the photometry is missing a
large number of sources. Because of the effects of extended
backgrounds and source confusion, the actual level of completeness of
the photometry depends on the location in the image (see
\S\ref{completeness}). Nevertheless, Figure \ref{photohist} shows that
below 31, 24, 80, 80, 400, and 40000 $\mu$Jy the point source
photometry for the 3.6 through 70 \um\ wavebands is clearly
incomplete.  These limits are very close to 10 times the 1-$\sigma$
noise in the images, except for the 3.6 and 4.5 $\mu$m wavebands where
that criterion would lead us to expect completeness limits of 7 and 11
$\mu$Jy respectively. The diminished effective ``sensitivity'' in
these wavebands is due to the effects of crowding and confusion, which
are most severe at the shortest wavelengths.

As in other {\em Spitzer} imaging observations, we have found that the
largest source of false stellar detections are the diffraction
artifacts associated with bright stars \citep[e.g.,][]{babbedge06}.
Wherever possible we have pruned them from the final catalog, either
by hand or by a two-step process in the photometry, where we use a
different PSF for bright stars. In the end, however, the catalog
suffers from some artifact contamination that is difficult to
estimate. These artificial sources are usually clustered near bright
stars, and are generally detected in just one waveband.

\subsection{Comparison to Previous Infrared Missions}
\label{irasmsx}

In this section we compare the fluxes of different point sources in
existing catalogs with our {\em Spitzer} photometry.  The SMC was
imaged by previous infrared satellites, including IRAS, MSX, and ISO.
There are numerous cataloged sources in the SMC, but the lower angular
resolution of most previous missions (especially IRAS) can make it
difficult to confidently identify them with sources seen in our data
or at other wavelengths. Also, the different instrument bandpasses
should be considered for a fully meaningful comparison of fluxes.  In
the following discussion we concentrate on the samples for which
source confusion and bandpass issues are minimized.

\subsubsection{MIPS Comparison}

Using the IRAS images, \citet{gb98} located 30 likely AGB stars in the
SMC.  Approximately one third of these sources are carbon stars, one
third are oxygen-rich AGB stars, and the remainder are presumed to be
other types of evolved stars.  The IRAS 25~$\mu$m bandpass is fairly
similar to the MIPS 24~$\mu$m bandpass, allowing us to compare the
flux densities we observe with \emph{Spitzer} to the earlier IRAS
observations. Nineteen of the stars discussed by \citet{gb98} lie
within the area covered by our MIPS 24~$\mu$m map, and 17 have good
fluxes in the IRAS and MIPS data.

We find that on average, the IRAS flux densities are 37\% higher than
the MIPS flux densities (23\% higher if we consider the median rather
than the mean), and for a few sources the difference is as
much as a factor of three.  Only two of the flux densities agree within
the quoted uncertainties given by \citet{gb98}.  \citet{trams99} also
found a similar discrepancy in comparing IRAS and ISO photometry for
sources fainter than 0.2 Jy. Some of these differences are certainly
due to variability \citep{w80}, which in dust-shrouded AGB stars can
amount up to a magnitude at mid-infrared (MIR) wavelengths
\citep{lebertre92,lebertre93,vanloon98}. Variability may introduce a
systematic offset in the sense of one data set consistently yielding
brighter measurements if that data has considerably lower
sensitivities and the sources in question are close to the detection
limit (in other words, they are only detected by IRAS in their high
state). Confusion within the large IRAS beam may also play a role,
although sources that are clearly isolated in the \emph{Spitzer}
observations show similar discrepancies between their IRAS and MIPS
fluxes, suggesting that confusion is not the main cause of the
disagreement.

The MSX Point Source Catalog \citep{msxps03} contains 9 very bright
($\gtrsim 1$~Jy) 21 \um\ sources in the SMC.  Eight of these lie
within our survey area, and seven are unsaturated. Two of these split
into multiple sources by MOPEX. Comparison of the MSX
21~$\mu$m flux densities with the MIPS 24~$\mu$m flux densities for
the remaining five sources shows that one of the sources agrees within
1-$\sigma$, and the remaining four are somewhat brighter ($13-45$~\%) in
the MIPS data. Overall, the average flux ratio between MSX and MIPS is
0.85, and the median is 0.88.  Given the substantial difference
between the MIPS and MSX bandpasses, and again allowing for
variability, it seems that the calibration of the 24 \um\
\emph{Spitzer} data is consistent with that of MSX.

\subsubsection{IRAC Comparison}

At 8 \um\ there are about 120 MSX sources in our field.  After
matching them to the IRAC catalog and removing the sources that are
flagged by MSX as being variable, confused, or low quality, 53 sources
remain. These sources are on average $\sim10\%$ brighter in our data
than in MSX.  This effect is entirely attributable to the fact that
the IRAC 8 \um\ bandpass is slightly bluer than that of MSX
($\lambda_{IRAC}\approx7.74$ \um\ compared to
$\lambda_{MSX}\approx8.28$ \um). A stellar blackbody in the
Rayleigh-Jeans limit (valid for photospheric temperatures
$T_{ph}>1000$~K) will be approximately $(8.28/7.74)^2\approx1.14$
times brighter in the IRAC waveband. Thus, the 8 \um\ IRAC point
source calibration is consistent with MSX.

There is also some overlap between the ISO Magellanic Clouds
Mini-Survey (Loup et al., in preparation) and the area mapped by our
survey.  The ISO data were taken in the LW2 band, which is centered at
6.7~\um\ and has a FWHM of 3.5~\um.  Unfortunately, this bandpass does
not match any of the IRAC bands very closely, but the average of IRAC
5.8~\um\ and 8.0~\um\ flux densities should provide a reasonable
approximation of the LW2 flux density.  We find 114 sources in the ISO
catalog that have both 5.8~\um\ and 8.0~\um\ fluxes, have no close
neighbors, are not confused, and have an ISO S/N of at least 10.
After discarding the 10 reddest sources ($[5.8]-[8.0]>0.51$), and
using the mean of the 5.8~\um\ and 8.0~\um\ bands, we find that the
IRAC flux densities are $\sim$5\% larger than the ISO measurements
with a scatter of $\sim$33\%.  Given the uncertainties involved in
comparing the different bandpasses, the calibration of our IRAC data
appears compatible with that of ISO.

\subsection{Photometric Completeness}
\label{completeness}

We used artificial star tests to estimate the completeness of the
photometry.  We added scaled versions of the measured PRF at random
positions in each image and re-ran the MOPEX photometry to see how
many of the artificial sources were recovered.  Because of the large
variations in both the surface density of point sources and the
brightness of extended emission regions across the SMC, the
completeness is a complex function of both flux and position.  In
order to determine the approximate size of these effects, we carried
out completeness tests in four separate fields that span the range of
surface densities found in the data.  Field 1 is centered at
$(\alpha_{2000},\delta_{2000})=(00\hh47\hm53\fs5,-72\degr22\arcmin46\arcsec)$
in the southwest part of the Bar, with extremely high stellar density
and significant extended emission. Field 2 is at
$(\alpha_{2000},\delta_{2000})=(00\hh54\hm33\fs7,-72\degr34\arcmin23\arcsec)$
near the middle of the Bar with a high stellar density. Field 3 is at
$(\alpha_{2000},\delta_{2000})=(01\hh01\hm54\fs1,-72\degr00\arcmin27\arcsec)$
near the top of the Bar with a moderate stellar density and some
extended emission. Finally, Field 4 is well out in the Wing at
$(\alpha_{2000},\delta_{2000})=(01\hh11\hm49\fs2,-73\degr25\arcmin08\arcsec)$,
in a low density area.  Each field is 768 pixels on a side, covering
an area of 104.9 arcmin$^{2}$ in the IRAC bands and 1065.4
arcmin$^{2}$ in the MIPS 24 \um\ mosaic.

We summarize in Table \ref{comptab} the results of the artificial star
tests.  The worst crowding occurs at 3.6~\um, and in the highest
density regions the photometry is only 90\% complete to $[3.6]=16$
(110.3~$\mu$Jy).  In the lowest density regions the 90\% completeness
limit is fainter than $[3.6]=17$.  At longer wavelengths, the crowding
declines dramatically as the stellar photospheres that comprise a
large majority of the detected sources fall off as $\lambda^{-2}$ and
the detector sensitivity rapidly decreases.  In the MIPS data,
incompleteness is largely caused by the presence of bright dust emission
rather than point source crowding (as there are no very crowded
regions).  Thus, the completeness results in the four fields scale as
the percentage of the field that is covered by bright extended
emission.

\begin{deluxetable}{crcccc}
\tablecolumns{6}
\tablewidth{0pt}
\tablecaption{Point Source Completeness Statistics\label{comptab}}
\tablehead{
\colhead{Magnitude} & \multicolumn{1}{c}{Flux} & \colhead{Field 1} & \colhead{Field 2} & \colhead{Field 3} & \colhead{Field 4}\\
& ($\mu$Jy) & (\%) & (\%) & (\%) & (\%)}
\startdata
\sidehead{3.6 \um}
15 & 277.1 & 97.3 & 99.3 & 99.0 & 99.7 \\
16 & 110.3 & 90.7 & 91.7 & 96.0 & 98.0 \\
17 &  43.9 & 79.0 & 83.2 & 91.0 & 95.2 \\
\sidehead{4.5 \um}
15 & 179.4 & 96.7 & 96.3 & 99.0 & 100.0 \\
16 & 71.4  & 91.0 & 92.0 & 97.3 & 99.7 \\
17 & 28.4  & 72.0 & 79.4 & 91.8 & 95.6 \\
\sidehead{5.8 \um}
14   & 286.1 & 98.7 & 99.0 & 98.3 & 99.7 \\
15   & 113.9 & 94.3 & 94.3 & 97.0 & 99.3 \\
15.5 & 71.9  & 78.2 & 88.0 & 95.0 & 96.8 \\
16   & 45.3  & 30.0 & 42.0 & 65.0 & 59.3 \\
\sidehead{8.0 \um}
14   & 158.5 & 94.7 & 97.3 & 96.0 & 98.3 \\
15   & 63.1  & 55.0 & 85.3 & 94.0 & 78.9 \\
15.5 & 39.8  & 14.8 & 34.0 & 54.2 & 30.4 \\
16   & 25.1  & 4.8  & 9.4  & 15.2 & 6.9  \\
\sidehead{24 \um}
9  & 1821.9 & 88.3 & 98.3 & 91.3 & 95.1 \\
10 & 725.3  & 78.0 & 92.7 & 86.9 & 93.1 \\
11 & 288.7  & 8.2  & 21.4 & 18.2 & 23.3 \\
\enddata
\end{deluxetable}

\subsection{Catalog Compilation}
\label{catcomp}

We searched for sources in each waveband using APEX and the procedure
outlined in \S\ref{photometry}. The catalogs for the individual
wavebands were then merged using positional tolerances of 1\arcsec\
for IRAC, and 3\arcsec\ and 6\arcsec\ for MIPS at 24 and 70 $\mu$m
respectively. Catalog assembly proceeded from the shortest to the
longest wavelength; the closest source within the aforementioned
positional tolerances was associated with a pre-existing source, or a
new entry was created if no cataloged source existed within the
tolerances. We derived the source position in the final merged catalog
of all the {\em Spitzer} bands using the signal-to-noise weighted
average of the detections in the IRAC bands, or adopting the the 24 or
70 $\mu$m positions (in that order) if no IRAC detections were available.

We proceeded to cross-identify the {\em Spitzer} sources with the OGLE
II catalog \citep{udalski98} and the SMC catalog of the Magellanic
Cloud Photometric Survey \citep[MCPS,][]{zaritsky02} in the optical,
and with the 2MASS catalog in the near infrared
\citep{skrutskie06}. The fluxes reported by OGLE II were preferred to
those of the MCPS \citep[see][]{zaritsky02}. The positional tolerance
used for cross-identification was 1\arcsec. The final catalog compiled
has over 400,000 point sources, and is available electronically at the
project URL\footnotemark[1]. To produce a uniform catalog we expressed
all fluxes in $\mu$Jy, using the following photometric zero points for
BVIJHK: 4060, 3723, 2459, 1594, 1024, 666.8 Jy
\citep{cohenopt03,cohen2mass03}. For the {\em Spitzer} wavebands we
use the zero points listed in Table \ref{surveyprop}.

\footnotetext[1]{\url{http://celestial.berkeley.edu/spitzer}}

\section{Discussion}
\label{discussion}

The S$^3$MC data provide a rich resource for studying many aspects of
the interstellar medium (ISM) and star formation activity in the
SMC. The 3.6 and 4.5 \um\ wavebands are mostly sensitive to stellar
photospheres and very hot circumstellar dust, with some contribution
from bound-free transitions, free-free radiation, and Brackett
$\alpha$ recombination emission in \hii\ regions. Although the 4.5
\um\ waveband is free from AIB contribution, the 3.29 \um\ emission
feature (thought to be due to C-H bond stretching in PAHs) contributes
to the 3.6 \um\ intensity near molecular clouds. In cool stars the 4.5
\um\ waveband includes the CO fundamental bandhead, which will appear
in absorption. Both the 5.8 and 8.0 \um\ wavebands show molecular
material in the ISM and circumstellar envelopes, as well as
increasingly faint stellar photospheres.  They encompass the 6.2 \um\
emission feature and the very bright emission complex at $7-9$ \um\
thought to be dominated by C-C stretching modes (with some
contribution from in-plane C-H bending) of the bonds in PAHs and very
small carbonaceous dust grains. The 24 \um\ waveband is sensitive to
continuum emission from PAHs and VSGs, which are stochastically heated
to temperatures of $\sim150$~K by the interstellar radiation field
(ISRF) or by the central star in warm circumstellar envelopes. Thus,
the extended 24 \um\ emission is bright in dusty regions near massive
stars and stellar clusters. In material that is shocked or exposed to
very hard radiation fields associated with peculiar objects, the 24
\um\ waveband may also have a contribution from spectral lines, such
as [\ion{O}{4}] at 25.89 \um. Emission in the 160 \um\ waveband arises
mostly from large dust grains in the ISM in equilibrium with
the radiation field, which typically heats them to temperatures of
several tens of degrees Kelvin. The 70 \um\ waveband, finally, is a
combination of VSG and large grain emission, in proportions that
depend on the relative abundances of the grain populations
\citep{desert90}.

\subsection{The SMC in the Mid and Far Infrared}

\begin{figure*}
\includegraphics[width=7.1in]{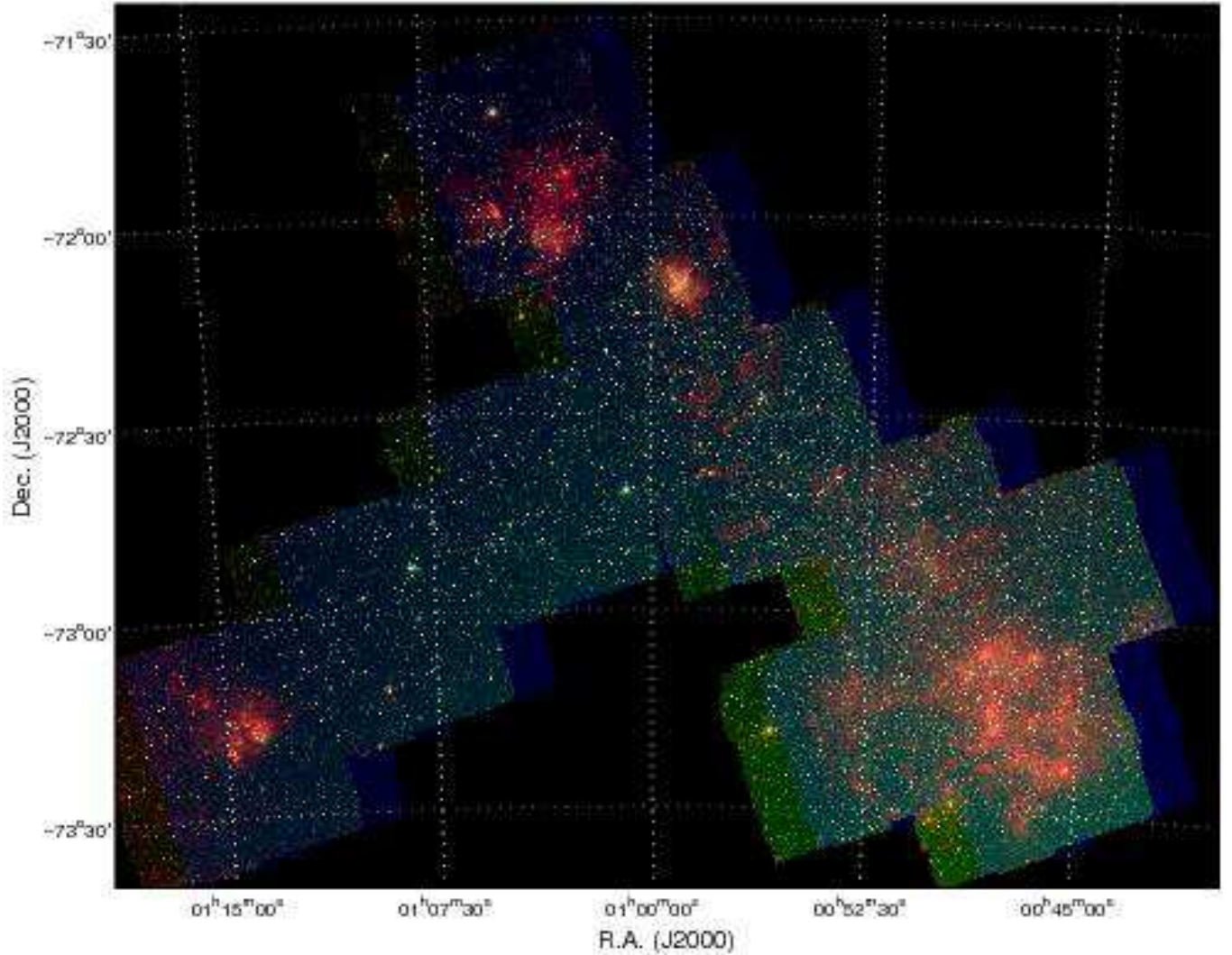} 
\figcaption{Color composite of the S$^3$MC mosaic for the IRAC 8.0, 4.5,
  and 3.6 \um\ wavebands, represented by the colors red, green, and
  blue. The prominent regions are N~83/N~84 at the SE tip of the Wing,
  N~76 in the north, N~66/NGC~346 near R.A.$=1\hh00\hm$ and
  Dec.$=-72\degr10\arcmin$, and the entire southwest region of the
  Bar, which hosts several prominent \ion{H}{2} regions such as
  N~22. Because of the arrangement of the different IRAC wavebands in
  the focal plane of {\em Spitzer}, the 3.6 \um\ data is displaced
  westward from the 4.5 and 8.0 \um\ data.\label{smc421}}
\end{figure*}

The {\em Spitzer} images of the SMC reveal an astounding degree of
detail. Figure \ref{smc421} is a color composite of the 8.0, 4.5, and
3.6 \um\ wavebands corresponding to the red, green, and blue colors.
The regions displaying bright red emission are associated with star
formation and known molecular clouds: N~66 (NGC~346), the region around
N~76, N~83/N~84, and the south-west tip of the Bar are all prominent at
8.0 \um. The south-west region is also the region of highest stellar
density, as evidenced by the bright diffuse blue background made up of
sources beyond {\em Spitzer's} confusion limit. Note that because the
3.6/5.8 \um\ and the 4.5/8.0 \um\ wavebands of IRAC occupy different
places in the {\em Spitzer} focal plane, the areas mapped by each pair
of channels do not completely overlap.

\begin{figure*}
\includegraphics[width=7.1in]{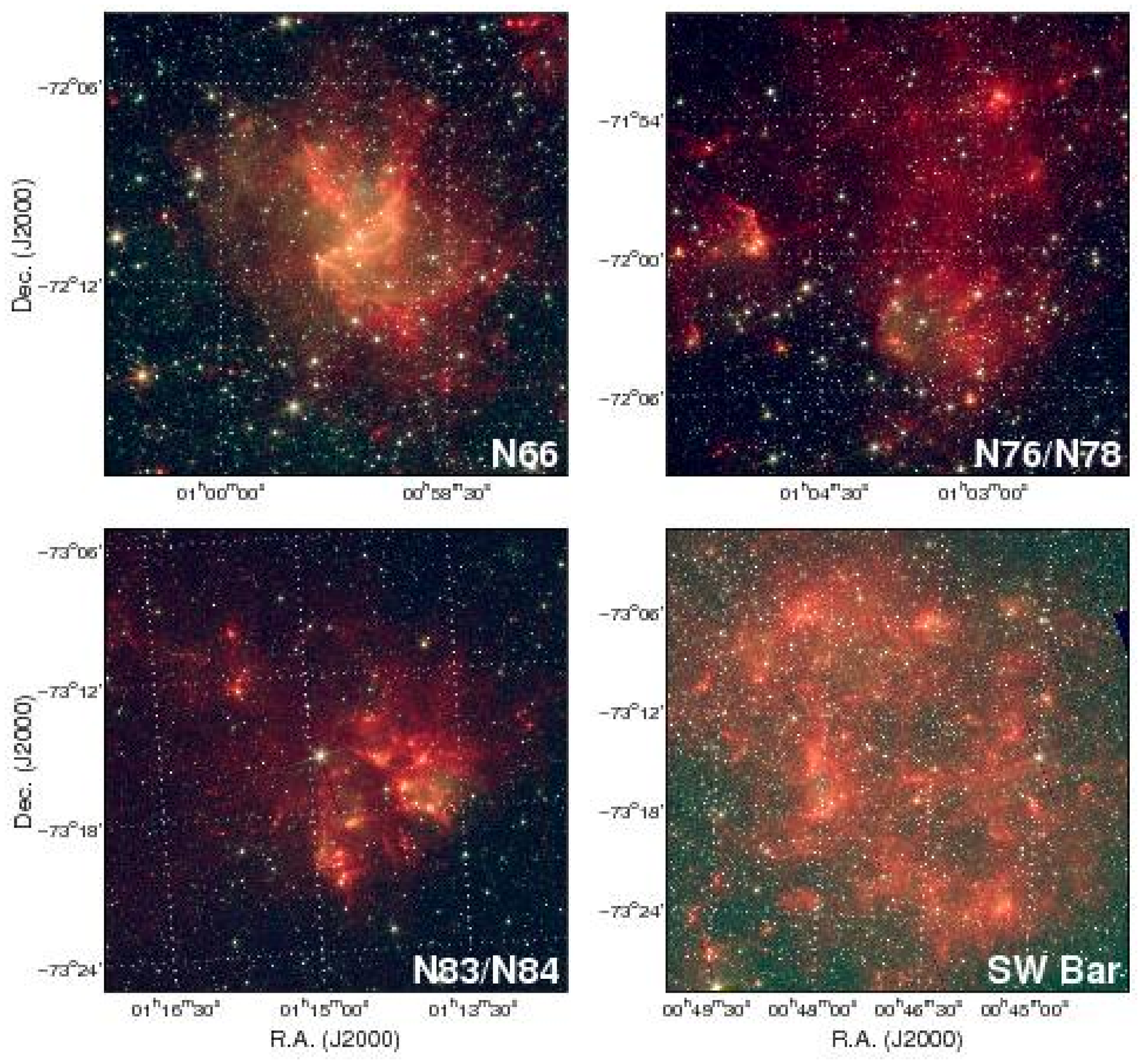}
\figcaption{Details of Figure \protect\ref{smc421} on four star forming
regions.  The sharp change in the MIR colors in the region surrounding
ionizing sources is particularly dramatic for N76 and N83/N84. These
regions are mostly devoid of 8 \um\ emission.\label{smc421det}}
\end{figure*}

Close inspection of Figure \ref{smc421} reveals that a dramatic change
of the colors of the MIR extended emission takes place within star
forming regions. Indeed, Figure \ref{smc421det} shows clearly that the
bright extended 8.0 \um\ emission frequently surrounds a well defined
cavity around one or more stars, inside which the emission is
bluer. This phenomenon is likely caused by the destruction of the
carriers of the 8.0 \um\ emission (nominally PAHs), which are
obliterated by the intense radiation field present in \hii\
regions. The transition between the regions of intense and faint 8
\um\ emission is often sharp, probably corresponding to the edges of
dense molecular material where shielding protects PAHs from
photodestruction by UV radiation.

\begin{deluxetable*}{ccccccc}
\tablecolumns{7}
\tablewidth{0pt}
\tablecaption{Resolved Flux Statistics\label{resflux}}
\tablehead{
\colhead{Waveband} & \multicolumn{1}{c}{Total Flux\tablenotemark{1}} & \multicolumn{1}{c}{Point sources} & \multicolumn{1}{c}{YSOs} & \multicolumn{1}{c}{Carbon stars} & \multicolumn{1}{c}{RSG/OAGB} & \multicolumn{1}{c}{Dusty B stars}\\
& \multicolumn{1}{c}{(Jy)}& \multicolumn{1}{c}{(Jy)}& \multicolumn{1}{c}{(\%)}& \multicolumn{1}{c}{(\%)}& \multicolumn{1}{c}{(\%)}& \multicolumn{1}{c}{(\%)}}
\startdata
3.6 \um & 163 & 95.9 & 0.2  & 13.6 & 3.5 & 0.0 \\
4.5 \um & 115 & 71.4 & 0.4  & 14.4 & 3.2 & 0.1 \\
5.8 \um & 85  & 71.1 & 1.4  & 17.2 & 3.2 & 0.1 \\
8.0 \um & 78  & 51.6 & 5.0  & 18.6 & 2.3 & 0.6 \\
24 \um  & 367 & 73.3 & 26.6 & 2.0  & 0.3 & 4.4 \\
70 \um  & 9451 & 347 & 19.7 & 0.1  & 0.0 & 3.4\tablenotemark{2} \\
\enddata
\tablenotetext{1}{Approximate total flux in the corresponding {\em
Spitzer} image after foreground subtraction. Note that the mapped area is
different in the various wavebands.}
\tablenotetext{2}{From only 6 objects detected. Probably misidentified 70 \um\
sources.}
\end{deluxetable*}

What percentage of the flux in the different wavebands is due to point
sources? Moreover, to help interpret the unresolved infrared
observations from distant metal-poor galaxies it is interesting to
estimate the contribution to the global infrared flux arising from
various types of point sources.  Table \ref{resflux} lists the fluxes
measured over the entire S$^3$MC images and compares it to the fluxes
in the different classes of point sources discussed in
\S\ref{starpops}. Considering the level of completeness in the
phometric catalog in the areas of high confusion (e.g., the SW Bar),
it is clear that most of the flux in the 3.6, 4.5, and 5.8 \um\
wavebands is due to point sources. At 8 \um, however, there is a
noticeable contribution from extended emission that becomes rapidly
dominant at the longer wavelengths of MIPS. Among the different types
of sources discussed in \S\ref{starpops}, the carbon stars stand out
prominently in the IRAC wavebands, while the YSOs are the largest
fraction of resolved flux for MIPS. The red supergiants and
oxygen-rich AGB stars, and the dusty early-B stars, although clearly
detected in the analysis, are not prominent contributors to any of the
{\em Spitzer} bands.

\subsection{PAHs in the SMC}

Among the recent {\em Spitzer} results is the disappearance of PAH
emission in galaxies of low metallicity
\citep{engelbracht05,hogg05,wu06,ohalloran06}.  The \fei/\ftf\ flux
ratio in galaxies changes dramatically at a metallicity of
$12+\log({\rm O/H})\sim8$ (the metallicity of the SMC), with low
metallicity objects exhibiting very low ratios (\fei/\ftf$\sim0.08$)
compared to those seen in solar-metallicity objects
(\fei/\ftf$\sim0.7$). This decrease in the 8 \um\ emission relative to
24 \um\ is frequently interpreted as the enhanced destruction or
reduced formation of PAHs in these environments
\citep{plante02,madden02,madden06}. PAH destruction by itself, however,
is probably not enough to account for the observed range in the
\fei/\ftf\ ratio, which may require also an enhanced hot dust
component in low-metallicity systems \citep{engelbracht05}.  Figures
\ref{smc421} and \ref{smc421det} clearly show that there are large
variations in the brightness of the 8 \um\ emission depending on the
local conditions. These images suggest on the scale of the entire SMC
an important conclusion reached by \citet{reach00} using observations
of a small, quiescent region: PAHs exist even in the low-metallicity
interstellar medium when the local environment allows it.

\begin{figure*}
\includegraphics[width=7.1in]{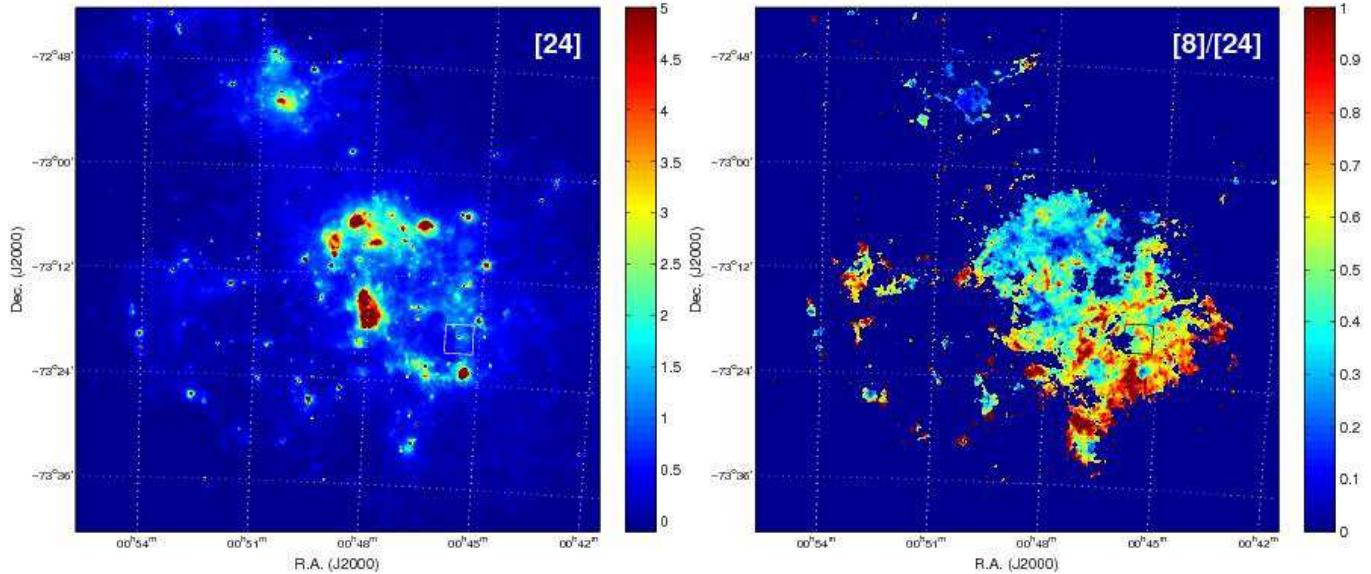}
\figcaption{The 24 \um\ image of the SW region of the SMC Bar at
  6\arcsec\ resolution, and the corresponding \fei/\ftf\ ratio, an
  estimator of PAH abundance. The units of the 24 \um\ intensity scale
  are MJy~sr$^{-1}$, and the display is saturated in the bright
  regions to emphasize the low level emission. Regions with surface
  brightness below 0.2 MJy sr$^{-1}$ in either band are masked out of
  the ratio.  There is an overall anticorrelation between bright 24
  \um\ emission and large \fei/\ftf\ ratios. The region studied by
  \protect\citet{reach00}, where PAHs were detected by ISO toward the
  molecular cloud SMCB1\#1, is outlined by a rectangle (white in the
  24 \um\ panel, black in the ratio panel). The area of high ratio
  within that region looks essentially identical to the 11.3 \um\ AIB
  map produced with ISO's CVF, suggesting that the \fei/\ftf\ ratio is
  an excellent diagnostic of PAH abundance.
\label{8vs24}}
\end{figure*}

We further explore the variation in the relative intensity of the 8
\um\ waveband by producing a ratio map of the SW region of the Bar
(Figure \ref{8vs24}). In order to produce this image we convolved the
8 \um\ image to a 6\arcsec\ resolution, subtracted the foregrounds
from each image (estimated from the modes of the respective histograms
to be 5.09 and 21.90 MJy~sr$^{-1}$ for the 8 and 24 \um\ images
respectively), and median filtered the resulting images to remove most
point sources. It is apparent that the \fei/\ftf\ ratio takes a large
range of values, with an overall gradient increasing toward the
southwest. This gradient has the opposite sense to the gradient in the
brightness of the 24 \um\ emission, an indicator of massive star
formation and intense ISRF. Extended regions of the SW Bar exhibit
$\fei/\ftf\gtrsim0.7$ typical of solar-metallicity objects, while
other areas generally associated with bright 24 \um\ continuum and
massive star formation have $\fei/\ftf\lesssim0.1$. Because there are
no large variations of metallicity across the SMC \citep{dufour84}, it
appears that the PAH abundance (at least as traced by the \fei/\ftf\
ratio) is regulated primarily by the local ISRF rather than
metallicity. This does not imply that metallicity plays no role on
setting the abundance of PAHs on galactic scales. In the context of
global measurements of galaxies, where the ISRF is a function of
metallicity, this observation suggests that photodestruction of PAHs,
not reduced formation, may be primarily responsible for the range of
values measured for the \fei/\ftf\ ratio. 

\begin{figure}
\plotone{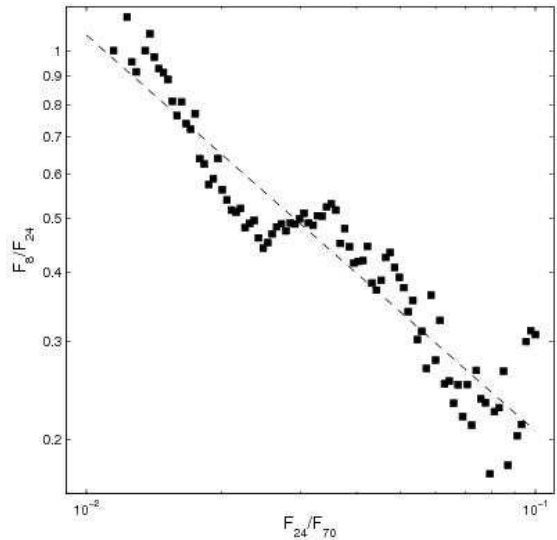} 
\figcaption{The \fei/\ftf\
ratio in the SW region of the SMC Bar as a function of the \ftf/\fst\
ratio. The black squares indicate the averages of the \fei/\ftf\ ratio
in bins of \ftf/\fst. There is a clear anticorrelation between the
\fei/\ftf\ and \ftf/\fst\ ratios, showing that regions with very blue
dust FIR colors are most likely underabundant in PAHs. The dashed line
shows a power-law fit to the data,
$\log[\fei/\ftf]\sim-0.7\log[\ftf/\fst]-1.4$\label{ratplot}}
\end{figure}

Figure \ref{ratplot} shows a quantification of the relationship
between this ratio and the dust FIR color $\ftf/\fst$. To obtain this
plot we convolved the 8 and 24 \um\ images of the SW Bar region to the
angular resolution of the 70 \um\ data, resampled the median-filtered
maps to obtain approximately one pixel per resolution element, and
removed the corresponding foregrounds from the individual images (in
addition to the 8 and 24 \um\ foregrounds mentioned above, we estimate
the 70 \um\ foreground to be $\sim6.5$ MJy~sr$^{-1}$). We considered
only data with $\fei>0.2$, $\ftf>0.2$, and $\fst>6$ MJy~sr$^{-1}$
after foreground removal. We computed the density of points in the
\fei/\ftf\ ratio versus the \ftf/\fst\ ratio plane, and calculated its
mean in \ftf/\fst\ bins. We find that, albeit with large scatter, the
trend for the \fei/\ftf\ ratio as a function of \ftf/\fst\ is
described by $\log[\fei/\ftf]\sim -0.7\log[\ftf/\fst]-1.4$. The
uncertainties in these parameters are considerable and due mostly to
uncertainties in the foreground substraction.  This shows that regions
with blue dust colors (thus presumably high radiation fields) tend to
be underluminous in 8 \um\ emission with respect to 24 \um, again
suggesting a link between the intensity of the local ISRF and the
presence of PAH emission.  The determination of the precise roles of
PAH destruction and enhanced 24 \um\ VSG emission in setting the
\fei/\ftf\ ratio will have to await upcoming spectroscopic
observations. About 55\% of the 8 \um\ flux and 22\% of the 24
\um\ flux in the SW Bar image are associated with regions where
$\fei/\ftf>0.5$. The overall ratio of the total 8 and 24 \um\ fluxes
in this region of the SMC is $\fei/\ftf\approx0.36$.  We defer the
detailed quantitative study of the extended emission of the SMC in the
FIR and MIR to a series of subsequent papers (Leroy et al., in prep.;
Sandstrom et al., in prep.; Bolatto et al., in prep.).

.

\subsection{Infrared Point Sources and Their Optical Counterparts}
\label{optical}

Perhaps one of the most interesting matters to pursue using the
S$^3$MC point source catalog is to investigate the classes of sources
detected by {\em Spitzer} in its different wavebands. The cross
identification against optical data is very valuable in this regard,
as the techniques of source classification based on the MIR colors and
magnitudes alone are currently in their infancy. The S$^3$MC catalog
provides unique deep photometric data on a large sample of stars
located at the same distance, with similar metallicities, and low
levels of foreground and background confusion. We expect that one of
the contributions of this dataset will be to provide a basis for
testing different source-identification schemes.

\begin{figure*}
\includegraphics[angle=-90,width=7.1in]{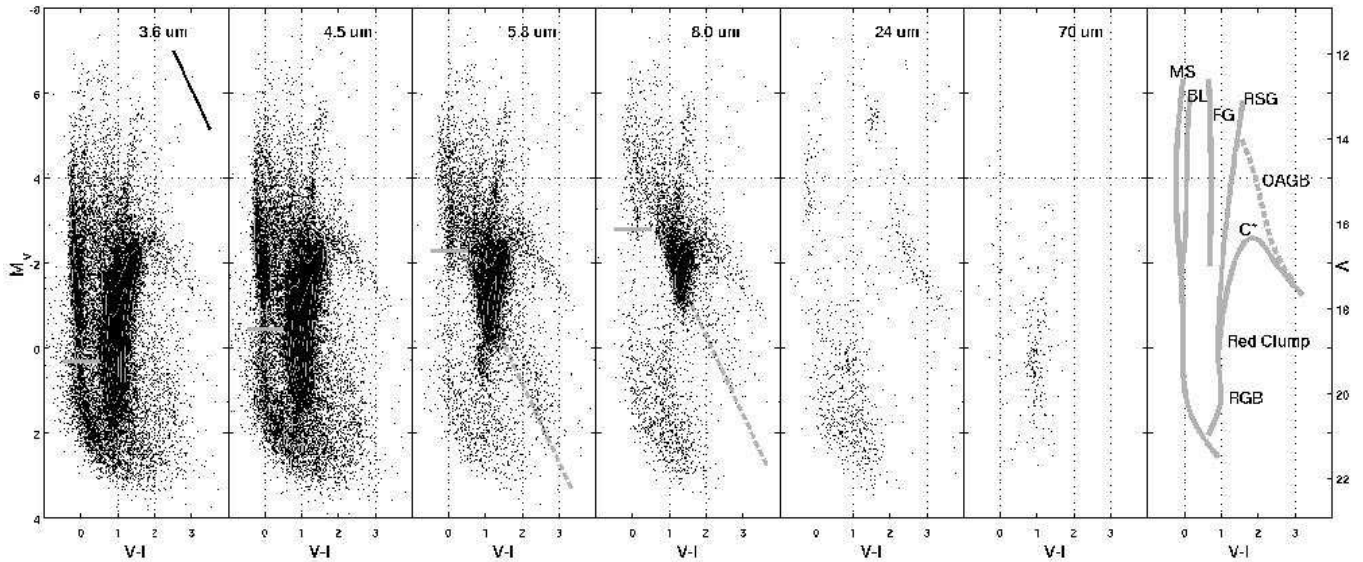}
\figcaption{Optical color-magnitude diagrams of the stars detected in the
different bands of {\em Spitzer} for which optical conterparts were
found in the OGLE II and the MCPS catalogs.  The overlaid gray
contours indicate density of sources. The assumed distance modulus
is $DM=18.93$, and the photometry has been corrected by the foreground
reddening and extinction toward the SMC ($A_V\sim0.12$,
$E(V-I)\sim0.09$).  The right side scale in the last panel correspond 
to apparent magnitudes. The gray bars indicate the expected 10-$\sigma$
detection thresholds for naked sources on the main sequence in the
different {\em Spitzer} wavebands, given the IRAC point source
sensitivities (c.f., Table \protect\ref{surveyprop}) and the known
color corrections. The vector in the upper right corner of the first
panel indicates the extinction correction for one magnitude of
reddening, assuming A$_V$/E($V-I$)$=1.85$ \citep{bouchet85}. The
dashed gray line in the 5.8 and 8.0 \um\ panels is parallel to the
extinction vector and indicates the boundary over which the density of
detected sources drastically increases. No naked photospheres should
be detectable in the MIPS data: detection at 24 and 70 $\mu$m requires
extensive reprocessing of the photospheric radiation by dust. All the
sources in the 70 $\mu$m diagram are consistent with
misidentifications (their distribution in the CMD plane is the same as
for the 3.6 or 4.5 $\mu$m sources). The rightmost panel sketches the
locii of the different branches with their identification: MS (main
sequence), BL (blue-loop, helium core-burning stars), FG (foreground), RSG
(red super giants), OAGB (oxygen-rich asymptotic giant branch), C*
(carbon stars), RGB (red giant branch). \label{VvsVI}}
\end{figure*}

Figure \ref{VvsVI} shows the optical color-magnitude diagrams (CMDs)
for sources detected in each {\em Spitzer} waveband at a
signal-to-noise greater than 10. Several populations are immediately
apparent: 1) the main sequence for stellar types earlier than A0, at
approximately zero $V-I$ color; 2) the blue-loop helium core-burning
stars, running parallel to the main sequence; 3) the red giant branch
and the red clump, where most of the {\em Spitzer} sources are found,
at $V-I\sim1$; 4) the asymptotic giant branch (AGB) and the red
supergiants (RSG); 5) the approximately horizontal branch
corresponding to evolved carbon stars, at $M_V\sim-2$ and $V-I>2$; and,
6) a fuzz of sources at low luminosities that consist of
misidentifications, stars in the Milky Way halo and thick disk,
unresolved background galaxies, and highly reddened sources. The
location of the blue-loop stars is both theoretically and
observationally linked to the metallicity of the intermediate-mass
stars that populate this feature; i.e., theoretical isochrones show a
metallicity dependence on the location of the blue loop in the CMD
\citep{girardi00,lejeune01}, and the observations of different
metallicity dwarf irregular galaxies confirm that the location of the
blue-loop stars varies with metallicity as predicted by isochrones
\citep[e.g.,][]{dohm-palmer97,dohm-palmer98}. In the SMC, the blue
loop stars are well separated from the main-sequence as expected for
their metallicities.  The foreground plume is nearly vertical at
$V-I=0.7$ whereas the RSG stars occupy a locus that is slightly
slanted at $V-I\sim1.1-1.2$ and $M_V\sim-3$ to almost $-6$.  The
foreground plume has been seen in other studies of the stellar
populations of Local Group dwarf galaxies
\citep[e.g.][]{mcconnachie05}.

As the sensitivity of our observations, indicated by the gray bar on
the main sequence, diminishes for the longer wavebands, we
progressively lose the fainter sources. The sensitivity across the
CMD, however, is clearly dependent on the $V-I$ color and follows a
trend, illustrated by the dashed gray lines in Figure \ref{VvsVI},
that is approximately parallel to the reddening vector illustrated in
the upper right corner of each panel. This reddening vector
corresponds to A$_V$/E($V-I$)$\approx1.85$, derived using SMC
measurements and assuming a Galactic E($V-I$)/E($V-J$)$\approx0.72$
\citep{bouchet85,riekelebofsky85}. The color dependence of the
sensitivity suggests that many sources in the highly reddened region
of the CMD could be intrinsically brighter stars located behind a few
magnitudes of extinction. The spatial distribution of these reddened
stars, however, is not strongly clustered. Such clustering would be
expected if they were located behind localized regions of extinction,
such as molecular clouds. Clearly, if the reddening of these stars is
caused by dust, this dust must be closely associated with the sources
themselves. Most of these sources, however, do not exhibit a strong
infrared excess, which makes this explanation unlikely. Alternatively,
these sources may be K and M dwarfs in the disk and halo of our own
Galaxy, at distances of less than 2 kpc. Their infrared colors
($J-H\sim0.6$, $H-K\sim0.2$) support this hypothesis. The slope in the
sensitivity would then be related to the slope of the lower main
sequence, weighted by the distance distribution of the sources. In
addition to these foreground interlopers, there is contamination by
background, unresolved galaxies, which can be seen throughout the SMC
at infrared wavelengths. We discuss the importance of these
contaminants in \S\ref{ysos}.

We do not expect to detect stellar photospheres in the MIPS wavebands.
Indeed, the drops in detector sensitivity and in stellar flux (which
approximately follows a $\lambda^{-2}$ Rayleigh-Jeans law), combine to
lower our detection limit by $\sim4.4$ mag at 24 $\mu$m with respect
to the 8.0 $\mu$m IRAC waveband. Thus, only the brightest red
supergiants and AGB stars are directly detectable at 24 \um. The other
sources found in the MIPS images must have a substantial infrared
excess from dusty envelopes or emission lines, or correspond to
misidentifications. The latter is probably the case with the optical
counterparts of most sources at 70 $\mu$m, which are distributed in
the CMD plane very much like a randomly drawn subsample of the
population detected at 3.6 or 4.5 $\mu$m.

Besides the bright end of the AGB and the RSG stars, there are three
other distinct populations of sources clearly detected at 24
$\mu$m. The bright end of the main sequence shows an overdensity of
sources with $V-I\approx-0.15$ and absolute $V$ magnitudes in the
range $-2.5>V>-4.5$, corresponding to stellar types B3 to O9. The
highly reddened side of the diagram shows the branch corresponding to
evolved carbon stars, with an overdensity of sources connecting it to
the red supergiants. Finally, the spread of sources at the bottom of
the diagram, which persists also throughout the IRAC bands, is not
entirely consistent with a randomly drawn subsample of the detections
at 3.6 or 4.5 $\mu$m. Although there is probably strong contamination,
if the bottom of the CMD were entirely composed of misidentified
sources we would expect it to look much more like the 70 $\mu$m
diagram, with a source density following closely the contours of the
CMDs for the lowest IRAC wavebands and featuring a prominent red
clump. It does not, thus a sizable fraction of these sources are
indeed bright MIR sources with very faint optical counterparts. It
turns out that many of them have very red MIR colors, and are
consistent with background galaxies and foreground M dwarfs. We will
investigate this matter further in \S\ref{starpops}.

The fact that the carbon star branch is prominent at 24 \um\ suggests
that it may be possible to use {\em Spitzer} observations to identify
these sources, which are common contaminants in extragalactic chemical
evolution studies. Carbon stars are difficult to analyze due to their
plethora of carbon molecular lines, which affect the continuum
placement in the analysis of the spectra. This problem is even more
accute in low-metallicity galaxies, where carbon stars are more
abundant \citep{blanco78,blanco80} and have stronger C$_2$H$_2$ bands
\citep{vanloon06b,sloan06}.  Recent work on dwarf spheroidal galaxies,
for example, has been based on detailed spectral analyses of RGB stars
\citep[e.g.,][]{shetrone03,tolstoy03,sadakane04}. The selection of the
target stars, however, is not always obvious and suffers from the
inclusion of carbon stars. In fact, because the studied galaxies are
far away the stars selected tend to be at the tip of the RGB, where
there is the greatest overlap and confusion potential with carbon
stars. In the aforementioned studies these contaminants amounted to
25\% of the original sample, and weeding them out required a large
time investment in an 8m class telescope. With infrared colors, target
selection can be significantly improved.  As shown by \citet{sloan06},
and also seen here, the carbon stars easily separate out from RGB
stars at infrared wavelengths, e.g., 3.6, 4.5, 5.8, 8.0, and even 24
\um.  Analysis of the old RGB stars in the dwarf irregular galaxies
may show similarities to the metal-poor halo of our Galaxy, and help
to explain the formation of our Galaxy, unlike the chemistry of the
metal-poor stars in the nearby dwarf spheroidal galaxies which do not
show chemical similarities with metal-poor stars in our Galaxy at all
\citep{venn04}.

\begin{figure*}
\includegraphics[angle=-90,width=7.1in]{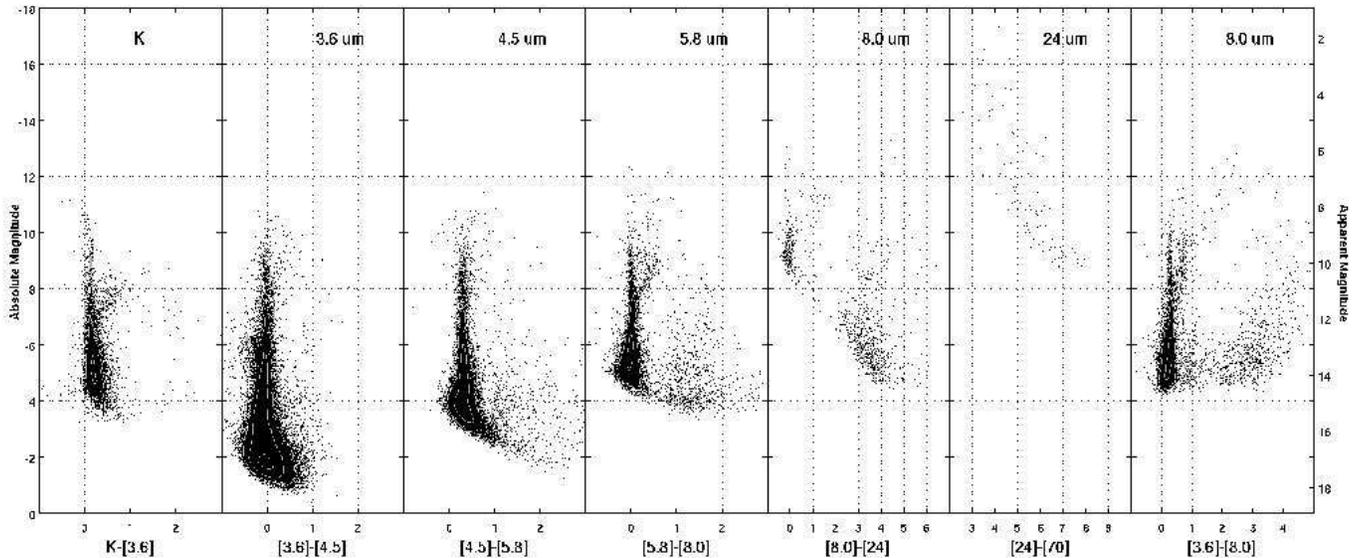}
\figcaption{{\em Spitzer} infrared color-magnitude diagrams of the SMC.  The
waveband corresponding to the ordinate axis is indicated in each
panel. The overlaid gray contours indicate densities of sources, and
the assumed distance modulus is $DM=18.93$.  Only sources detected with
signal-to-noise greater than 10 in the wavebands used in each CMD are
ploted. The right side scale on the last panel indicates apparent magnitude.
\label{cmdIR}}
\end{figure*}

\begin{figure*}
\includegraphics[angle=-90,width=7.1in]{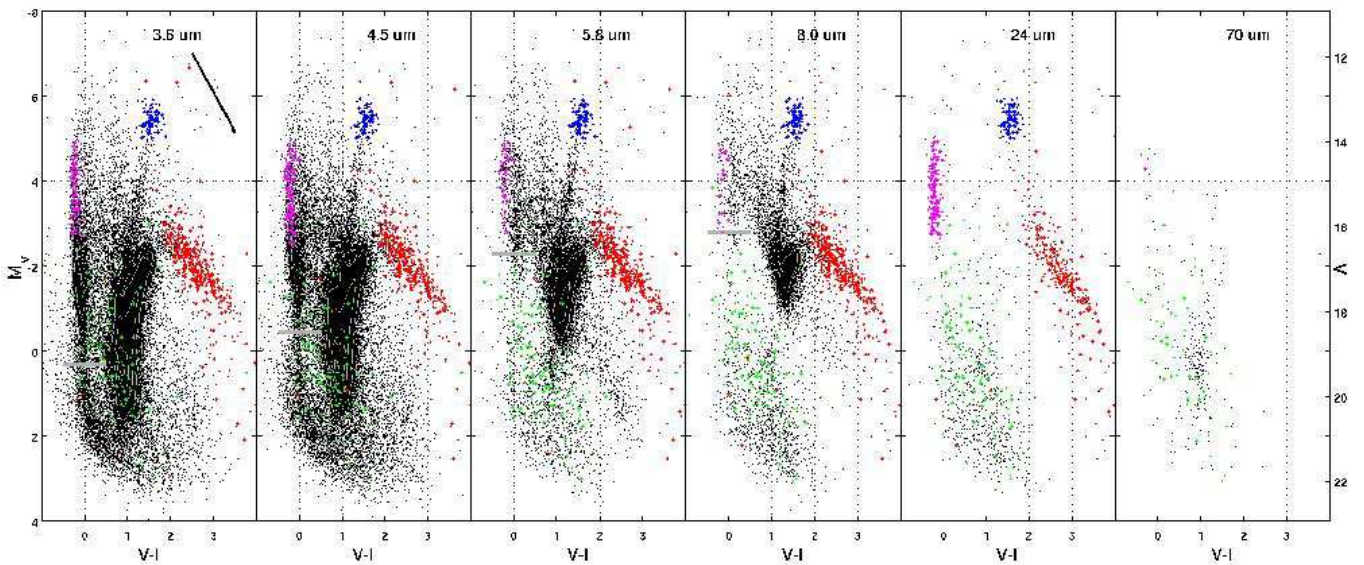} \figcaption{Same as
Figure \protect\ref{VvsVI}, now with different populations selected by
the magnitude and color cuts defined in the main text highlighted. The
green, red, blue, and magenta symbols correspond to populations number
1 (YSOs), 2 (carbon stars), 3 (RSG and OAGB stars), and 4 (B stars
with 24 \um\ emission) respectively. Most of the stars in population
1, sources with strong 8.0 $\mu$m excess, have only very faint optical
counterparts. Sources in population 2, which splits off the bright end
of the MIR main sequence, are clearly identified with the branch of
carbon stars. \label{VvsVIpop}}
\end{figure*}

\subsection{Mid and Far IR Color-Magnitude Diagrams}

Figure \ref{cmdIR} shows the color-magnitude diagrams in different
combinations of the {\em Spitzer} wavebands. The progression from left
to right shows the absolute magnitude in one band plotted against the
color with respect to the next {\em Spitzer} waveband. The rightmost
panel shows the 8 $\mu$m magnitude versus the $[3.6]-[8.0]$ color, a
CMD modeled by \citet{whitney03}.

Several features are apparent in these plots.  Since most stars emit
like blackbodies in the infrared, they pile on a vertical line near
zero color constituting a MIR ``main sequence''. At longer wavelengths
there is an increasing number of objects that appear as a red plume of
sources, which becomes progressively more detached from this main
sequence. In several of these CMDs there is also a number of sources
in a branch off the main sequence toward redder colors at high
luminosities. At the longer wavebands the progressive loss of
sensitivity results in a diminishing number of sources detected on the
main sequence, until in the 24 $\mu$m vs. $[24]-[70]$ CMD most sources
have a color redder than 3, implying a ratio of fluxes
$F_{70}/F_{24}\gtrsim1.8$ and a spectral energy density rapidly rising
toward the far infrared.

\subsection{Stellar Populations}
\label{starpops}

\begin{figure*}
\includegraphics[angle=-90,width=7.1in]{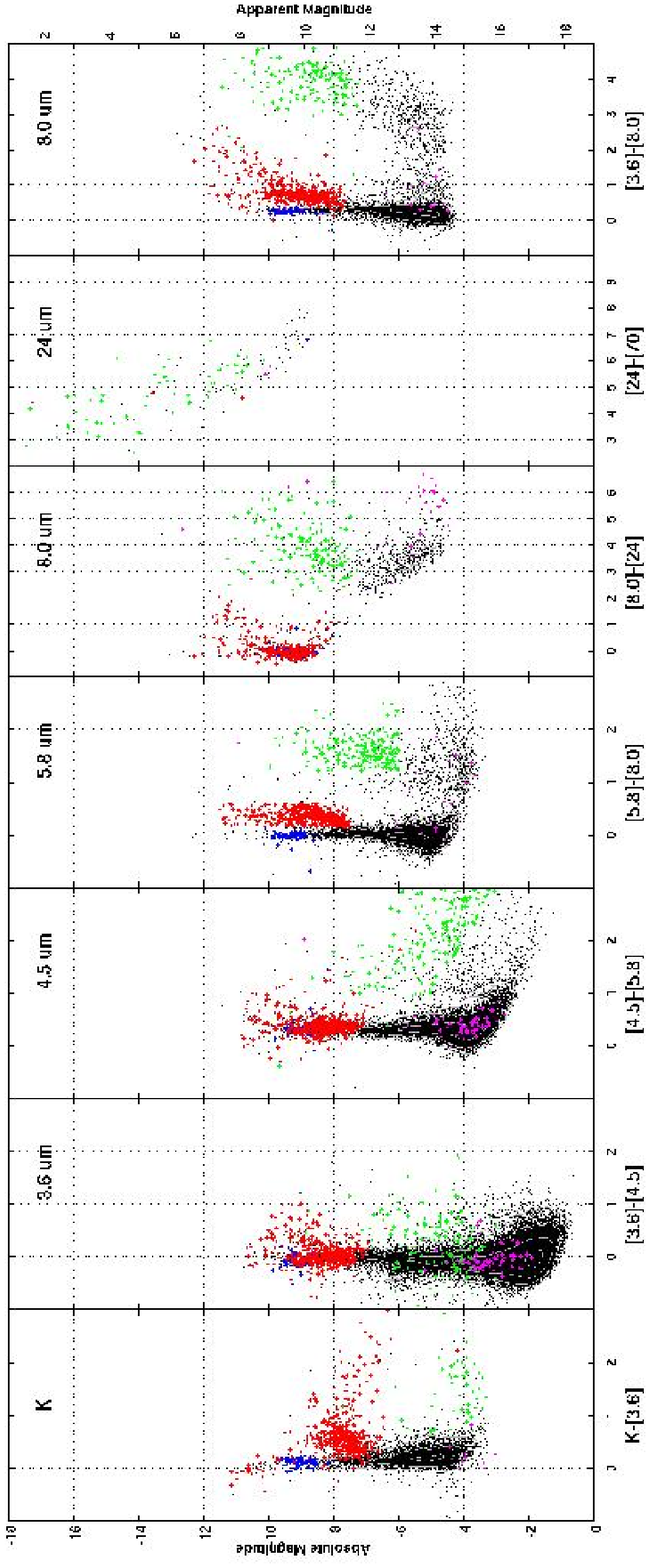} \figcaption{Infrared
color-magnitude diagrams, as in Figure \protect\ref{cmdIR}, with the
different populations identified using the color scheme of Figure
\protect\ref{VvsVIpop}.\label{cmdIRpop}}
\end{figure*}

To gain insight into the nature of the different stellar populations
detected, we used color and flux selection criteria in the CMDs to
pull out particular subsets of sources (Figures \ref{VvsVIpop} and
\ref{cmdIRpop}). We defined four groups of sources, which we discuss
below. Sources in groups 1 to 3 are a substantial fraction of the
brightest sources seen in our {\em Spitzer} data ($\gtrsim50\%$,
Figure \ref{photohist}).

\begin{enumerate}
\item Sources with
$[5.8]-[8.0]>1.2$, and $-10<M_{5.8}<-6$. We also require that these
sources be detected with $S/N>10$ at these wavebands and one
neighboring waveband (either 4.5 or 24 \um).  About 280 sources fall
into this category. They are representative of the highly reddened
plume present in the 5.8 $\mu$m CMD.

\item Sources with $0.16<[5.8]-[8.0]<0.6$ and $-12<M_{5.8}<-7.5$ and
$S/N>10$.  About 660 sources fall into this class. These correspond to
the branch that splits off the upper main sequence in the MIR. These
sources are identified by their optical CMD as carbon stars.

\item Sources detected with $S/N>10$ at 24 $\mu$m and with optical colors
$1.2<V-I<1.9$ and absolute magnitude $-6<M_V<-5$. About 130 sources belong to
this category. These represent the red supergiants and the bright end
of the OAGB (i.e., oxygen-rich evolved stars) detected at 24 $\mu$m. 

\item Sources detected with $S/N>10$ at 24 $\mu$m and at $3.6$ or
$4.5$ $\mu$m, with optical colors $-0.4<V-I<0$ and absolute magnitude
$-5<M_V<-2.5$. These sources correspond to the population of early-B
stars (B3 to O9) detected at 24 $\mu$m; about 190 (5\%) of the stars
that fulfill these optical color criteria have 24 $\mu$m
counterparts. Inspection of the individual SEDs to cull sources with
poor or likely misidentified photometry reduces the sample to
$\sim120$ sources.
\end{enumerate}

These populations have been highlighted in the optical CMDs in Figure
\ref{VvsVIpop}. This plot shows that the stars with large 8.0 $\mu$m
excess selected in the MIR CMDs (group 1 defined above) have only very
faint optical counterparts, as they are almost all located in the
spread of sources at the bottom of the diagram. Many of these
counterparts may in fact be misidentifications.  The bright stars with
small 8.0 $\mu$m excess (group 2) are clearly identified with the
branch of carbon stars in the optical. Bright carbon stars have been
studied using {\em Spitzer} spectroscopy by \citet{sloan06}. They find
that oxygen-rich and carbon-rich AGB can be split into two sequences
using $J-K$ and $[8]-[24]$ colors, with the oxygen-rich and
carbon-rich sources forming two sequences around $1<J-K<2$ and
$[8]-[24]\sim1-3$, and $J-K>2$ and $[8]-[24]\sim0.8$ respectively
\citep[see also][]{blum06}. We clearly see both sequences in Figure
\ref{colcol}d. The carbon stars, as well as the red supergiant stars
selected in group 3, are not preferentially distributed along the SMC
Bar but show a much smoother spatial distribution (Figure
\ref{smcpop}b,c). In particular, the distribution of the MIR-selected
carbon stars is compatible with the spheroidal shape found by other
studies of the older SMC stellar populations
\citep[e.g.,][]{cioni00,zaritsky00}.

Identification of these populations in the CMDs in Figure
\ref{cmdIRpop} shows some of their MIR characteristics. The red
supergiants, for example, are also most of the brightest stars in the
MIR. Many of the rest of the brightest stars are carbon stars, which
form a distinct branch off the main sequence not only in the
$[5.8]-[8.0]$ color, but also in the $K-[3.6]$ CMD.  The B stars with
24 $\mu$m excess cluster near the bottom of the main sequence in most
of these CMDs due to their large bolometric corrections at IR
wavelengths, although they tend to be under-represented at the longest
wavelengths because very few are detected at 5.8 or 70 $\mu$m.

\subsubsection{A Sample of Young Stellar Objects in the SMC}
\label{ysos}

\begin{figure*}
\includegraphics[width=7.1in]{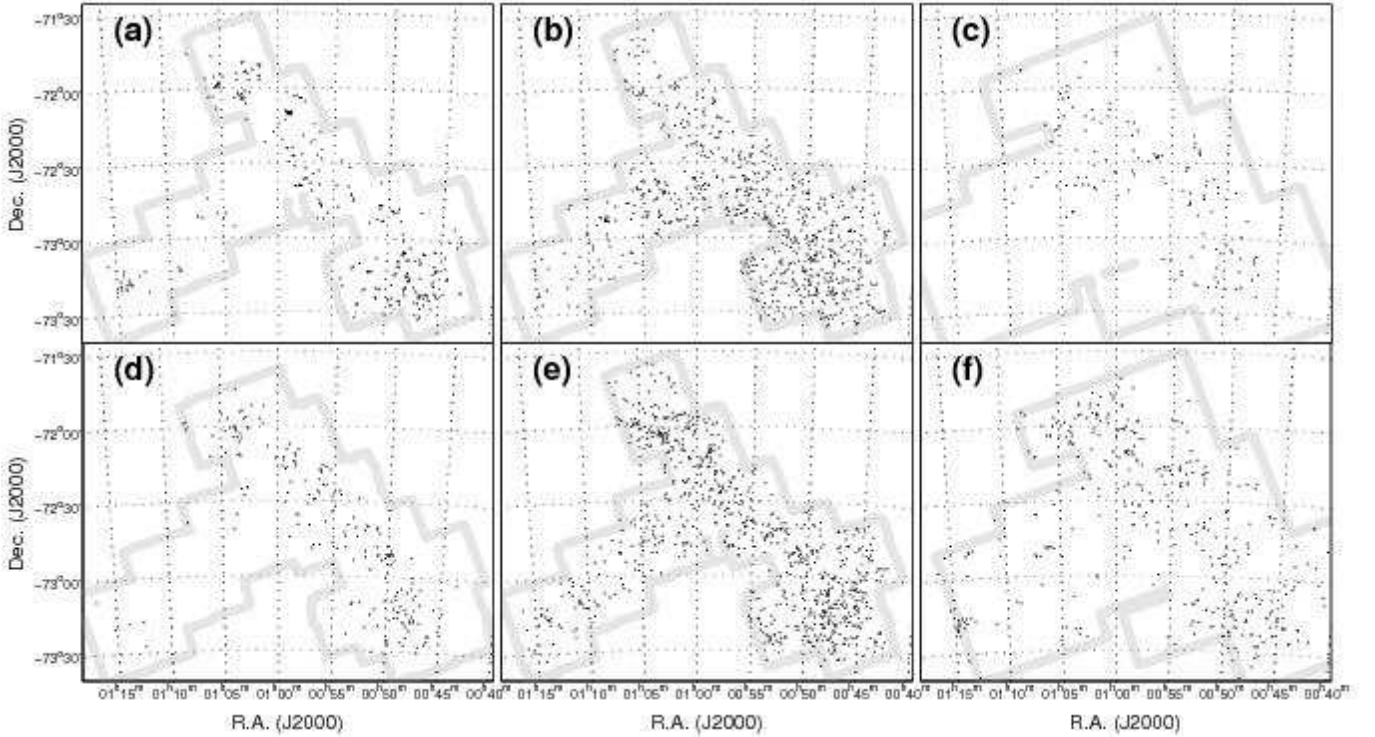}
\figcaption{Distribution throughout the SMC of the different populations
of sources selected using the color-magnitude cuts described in the
main text: (a) YSOs. (b) Carbon stars. (c) Red Supergiants and bright
end of the AGB. (d) Early B stars with 24 $\mu$m emission. (e) Faint
and red MIR sources, corresponding to YSOs heavily contaminated by
background galaxies. (f) Highly significant 70 \um\ sources without 24
\um\ counterparts (S/N$>25$). The gray outline indicates the
approximate region covered by the combination of wavebands used to
identify the different populations.\label{smcpop}}
\end{figure*}

\begin{figure}
\plotone{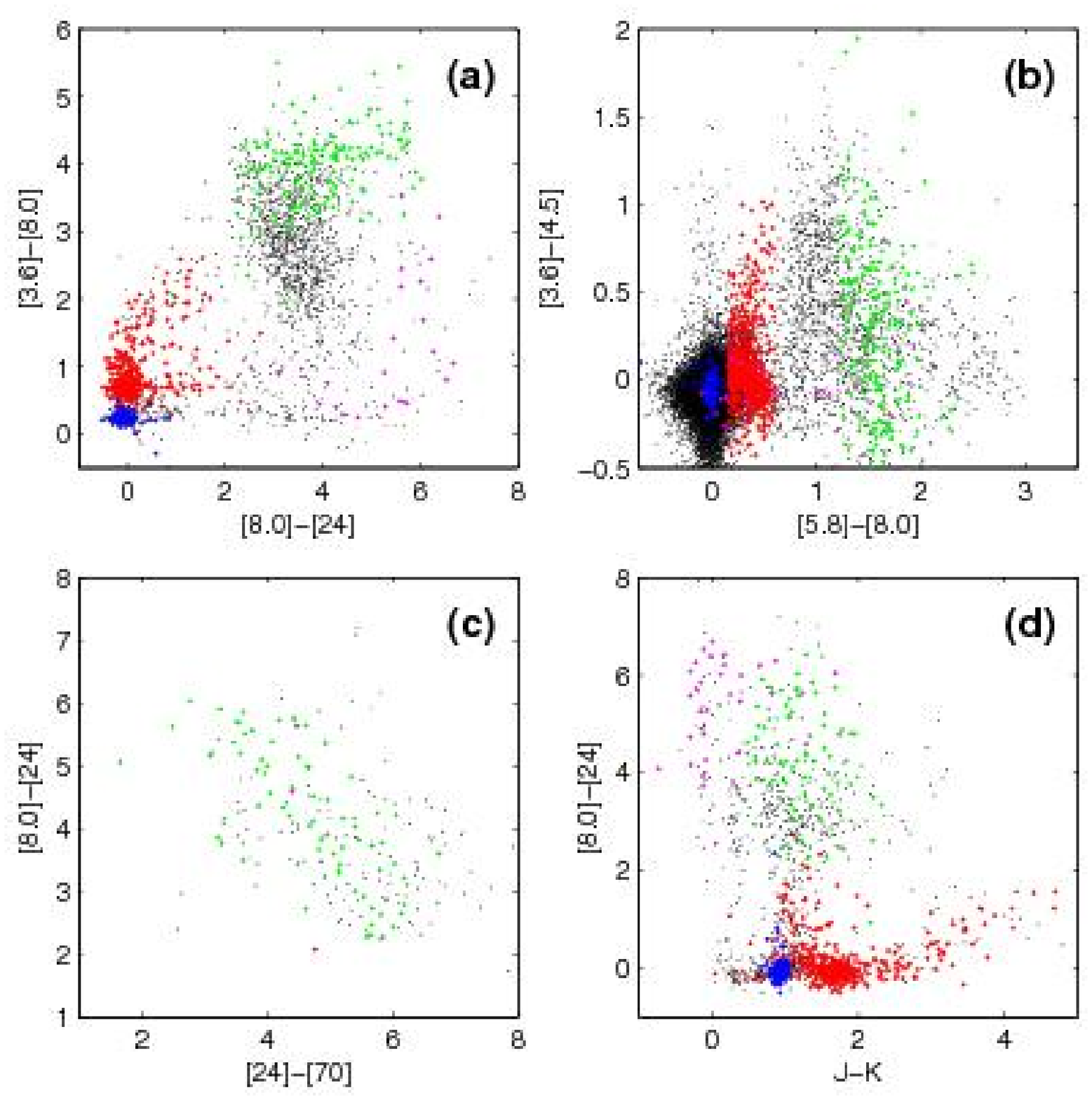} \figcaption{Color-color plots in the {\em Spitzer}
and 2MASS wavebands. The color of the symbols is identical to that
employed in Figures \ref{VvsVIpop} and \ref{cmdIRpop}.  Panels a, b,
and c, show plots similar to those modeled by \citet{whitney03} for a
variety of sources. The group 1 sources (green symbols) fall into the
locii of Class 0 to Class II YSOs. Panel d, a plot similar to Figure 1
of \citet{sloan06}, clearly shows the two sequences of silicate and
carbon rich dusty evolved stars (red symbols).\label{colcol}}
\end{figure}

The highly reddened 8.0 $\mu$m sources in group 1 constitute many of
the sources detected at both 24 and 70 $\mu$m. Their spatial
distibution (Figure \ref{smcpop}a) is highly clustered around known
molecular clouds and star forming regions, suggesting that many of
these sources are still embedded young stellar objects (YSOs), with a
secondary contribution from unresolved extragalactic background
sources. Further support for this hypothesis is lent by the location
of these sources in the [8.0] vs. $[3.6]-[8.0]$ CMD, as well as in the
color-color plots in Figure \ref{colcol}. These combinations of colors
have been modeled by \citet{whitney03} for different objects,
including YSOs. The highly reddened 8.0 $\mu$m sources occupy the
space in those plots that is predicted for late Class 0 to Class II
YSOs with different geometrical parameters. About 280 candidate YSOs
fulfill our color-magnitude criteria, and they are compiled in Table
4 (published in its entirety in the electronic
edition of the journal). The color-magnitude selection criteria used
to identify these sources have been chosen to minimize the
contamination of the sample, and limit us to the brightest YSOs in the
SMC; clearly there is a large population of fainter YSOs detected in
this survey as well, but it is highly contaminated by background
galaxies (Figure \ref{smcpop}e).  By modeling the SEDs of individual
sources rather than relying only on color selection it is possible to
identify a larger samples of YSOs in these data. Simon et al. (in
prep.) successfully use this technique to identify over 200 YSOs with
stellar masses as low as 1.5~M$_{\odot}$ in the N~66 region alone.

Missing from Figure \ref{colcol} are the early Class 0 YSOs, which are
predicted to have extreme $[24]-[70]$ colors $\gtrsim12$. Our
sensitivity at 24 $\mu$m would not, for the most part, allow us to
detect such sources. Thus, some of our 70 $\mu$m detections without 24
$\mu$m counterparts may be early Class 0 YSOs (Figure \ref{smcpop}f).
There are approximately 370 highly significant ($S/N\gtrsim25$) 70
\um\ detections without 24 \um\ counterparts in our photometry, with a
total flux $F_{70}\sim127$ Jy (i.e., almost twice the total flux in
our sample of YSOs in Table \ref{ysos} YSOs as noted in Table
\ref{resflux}).  Because of the limited angular resolution at 70
$\mu$m, the image artifacts, and the bright extended emission result
in catalog reliability and confusion issues, however, we defer such
a study to a separate paper (Simon et al., in prep.).

To quantify the degree of contamination by background galaxies and
foreground thick disk interlopers we have used the SWIRE catalog of
the ELAIS-N1 field by \citet{surace04}.  This field is conveniently
located at the same Galactic latitude as the SMC but in the northern
hemisphere, and at a Galactic longitude $l\sim+90\degr$ comparable to
that of the SMC ($l\sim-50\degr$). Applying exactly the same color and
magnitude criteria we used to find YSOs in the S$^3$MC catalog to the
SWIRE catalog we find 70 sources in the entire 8.7 square degree
ELAIS-N1 field \citep{lonsdale03}. Scaling this number to the 5.8 and
8.0 \um\ overlap area in the S$^3$MC mosaic (2.45 square degrees), we
expect $\sim20$ contaminants in our sample of $\sim280$ candidates
($\sim7\%$). The magnitude and color cuts employed are designed to
minimize the degree of contamination: pushing below $M_{5.8}=-6$ rapidly
increases the number of background galaxies included. Reducing the
faint magnitude cut to $M_{5.8}=-5$ results in $\sim110$ interlopers out
of 533 sources ($\sim21\%$).  Moving the limit in $[5.8]-[8.0]$ to
bluer colors, although more benign than including fainter sources,
results also in a higher rate of contaminants.  Using
$[5.8]-[8.0]=0.8$ (1.0) yields 40 (27) interlopers in a total of 386
(340) sources, or $\sim10\%$ (8\%) contamination. Figure \ref{smcpop}e
shows the spatial distribution of sources that have the same
$[5.8]-[8.0]$ colors as our YSO candidate sample but that are fainter
than $M_{5.8}=-5$. Although there is an concentration of such sources
along the Bar and in known star-forming regions, they are much more
uniformly distributed than our YSO sample, pointing to heavy
contamination by sources outside the SMC.


\subsubsection{A Population of Dusty Early B Stars in the SMC}

\begin{figure}
\plotone{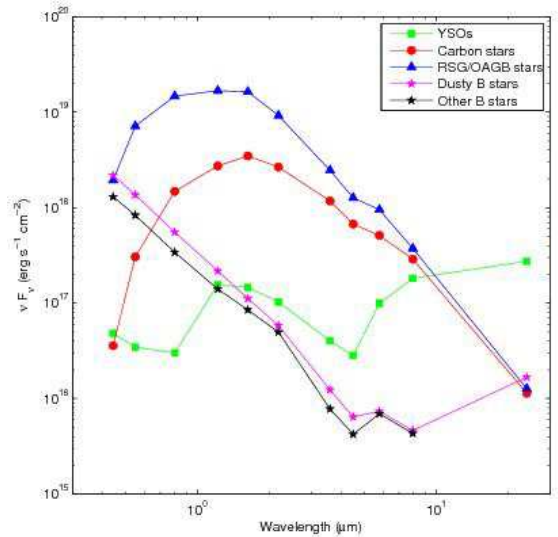} \figcaption{Average spectral energy distributions
of the different populations described in the main text. The SEDs were
obtained as the median at each wavelength over the entire distribution
of objects.  The drop in the SED at 4.5 \um\ seen in carbon stars and
RSG and OAGB stars is probably due to the CO fundamental bandhead at
4.7 \um\ seen in absorption.
\label{seds}}
\end{figure}

Figure \ref{VvsVIpop} shows that the early B stars detected at 24
$\mu$m, defined as group 4 above, have an extremely interesting
spectral energy distribution.  They gradually disappear in the 5.8 and
8.0 $\mu$m wavebands, then strongly reappear at 24 $\mu$m, and there
are virtually no detections at 70 $\mu$m.  These 24 $\mu$m-bright
stars are not conspicuous at any other wavelength. Their typical
absolute magnitudes in any of the IRAC bands are $M_{IRAC}\sim-4$,
while their mean 24 $\mu$m absolute magnitude is
$M_{24}\sim-10$. Their mean SED (Figure \ref{seds}) shows that the
typical 24 \um\ excess over the photospheric flux is very large,
$F_{24}/F_{photo}\sim200$, otherwise we would not have detected
them. Their $K-[24]$ color, an excess indicator widely used because of
its independence from the stellar temperature, is extremely large,
with a typical value $K-[24]\sim6.5$.

The sharp rise in the SED of these stars at 24 $\mu$m can be explained
by emission from $\sim0.05-1.4$ M$_{\earth}$ of dust at a temperature
of T$_{dust}\sim150-200$~K, or perhaps by an emission line present in
the 24 $\mu$m window. It is difficult to imagine what such an emission
line could be, since the usual candidates such as [\ion{O}{4}] at
25.89 \um\ or [\ion{Fe}{2}] at 24.52 \um\ are not prominent in normal
stars.  If dust is the cause of their detection at 24 $\mu$m, however,
a particular geometry is required because these stars are not
appreciably reddened in the optical. Their intrinsic colors are
$B-V\approx-0.20\pm0.12$ and $V-I\approx-0.25\pm0.07$ (after removing
the line-of-sight reddening to the SMC), while the photospheric colors
of B0 stars of metallicity $Z=0.1-0.3 Z_\odot$ are $B-V=-0.27$ and
$V-I=-0.36$. The preferred geometry would be a face-on disk if the
material is close to the star, a tenuous envelope, or a thin disk with
a large central gap seen at arbitrary viewing angles.  It could be
argued that, since these stars were selected from their optical
colors, the fact that they are only slightly reddened is not
surprising. The FWHM of the 24 \um\ emitting B star sequence seen in
Figure \ref{VvsVI}, however, is only 0.08~mag in $V-I$.  This is
completely consistent with the FWHM of the normal, main sequence B
stars (0.07~mag).  The 24~\um-bright B stars therefore form a sequence
as sharp as the main sequence with no sign of additional spread due to
intrinsic reddening. Both this fact and Figure \ref{VvsVI} show that
there is no evidence of a population of more reddened but otherwise
similar sources. This suggests that the dusty material is arranged in
a very tenuous envelope or in a thin disk with a large central gap, so
that viewing angle does not play a role in the selection of these
stars. Given the very large luminosity of the central source, material
at T$_{dust}\sim150$~K near an early-B star should be located at
distances of $\sim350$~AU or more \citep{backman93}.

What are these ``24 \um\ excess'' early-B stars?  Searches for these
objects in the SIMBAD and VizieR databases indicate that out of
approximately 190 stars in this group, 10 are eclipsing binaries
\citep[out of 1351 such objects found by OGLE in the
SMC;][]{wyrzykowski04}, and 12 are emission line stars
\citep{meyssonnier93}. In these aspects, this population does not seem
peculiar.  These sources are present in the SMC Bar and the tip of the
Wing (Figure \ref{smcpop}d), in regions of recent star
formation. Their spatial distribution is, in fact, similar to that of
the YSO candidates although not as strongly clustered, which is
consistent with expectations for $\lesssim10$~Myr old stars.  We consider
four possible explanations for the origin of these objects and their
24 \um\ emission: 1) an accretion disk remnant, 2) free-free emission
associated with emission line stars, 3) a circumstellar debris disk,
or 4) nearby cirrus emission unrelated to the stars themselves. We
briefly discuss these possibilities below, but an in-depth discussion
of these sources is deferred to a forthcoming paper (Bolatto et al.,
in prep.).

Far infrared excess in stars is frequently a sign of stellar youth.
These dusty B stars may be in the process of shedding their accretion
remnants through photoevaporation.  These early-B stars exhibit small
ratios of IR excess luminosity to stellar luminosity,
$10^{-4}<L_{IR}/L_*<10^{-2}$, rather than the much larger ratios
observed in young stellar objects, $L_{IR}/L_*>10^{-1}$. This fact,
and their optical colors, suggest that their 24 \um\ emission may be
due to a remnant accretion disk or envelope with a large central gap,
and that we may be observing them in the very late stages of the
accretion process.  Disk dispersal timescales, and in particular
the timescale for removing the outer disk, are not well known for
massive stars.  Models of less massive stars, however, suggest that
the outer disk clearing time is dominated by photoevaporation, and
that its dispersal occurs very quickly after the inner disk is cleared
\citep{alexander06a,alexander06b}.  This phase in the lives of stars
would be, consequently, very short, which suggests its observation is
improbable.

Although Milky Way emission line stars frequently have IR excesses, it
seems unlikely that the 24 \um\ emission observed in our SMC objects
is explained by the various types of Be activity.  A type of infrared
excess associated with this activity is observed in Galactic Herbig
Ae/Be stars, which are young, massive stars that may host
circumstellar disks or envelopes that remain from the protostellar
accretion phase and give rise to IR emission
\citep{hillenbrand92}. Classical Be stars, on the other hand, are
older objects whose IR excess comes from free-free emission associated
with their stellar winds and circumstellar disks. Be stars in the
Milky Way tend to be later spectral types, but there is some evidence
pointing to a large abundance of very early spectral types in the
Magellanic Cloud \citep{grebel93,wisniewski06}. We already mentioned
that only a few of these 24 \um\ emitting early-B stars are identified
as Be stars in existing catalogs. Be envelopes are hot and close to
the star, showing an infrared excess at $1-2$ \um\
\citep[e.g.,][]{waters98}. Most, albeit not all, of our objects do
have a small excess shortwards of 24 \um. At 4.5 \um\ the measured
flux is 75\% over the photosphere predicted by a Kurucz model tied to
the observed V magnitude, and only a handful of these stars show an
excess over a factor of 2. By comparison, in the prototypical Herbig
Ae/Be star AB Aur the observed emission exceeds the photospheric
emission by more than an order of magnitude at 4.5 \um, and similarly
large excesses are observed in many classical Be stars
\citep[e.g.,][]{miroshnichenko03}. Furthermore, the mean SED of our 24
\um\ emitting B stars shortward of 24 \um\ is indistinguishable from
the mean SED of B stars with the same $V$ and $V-I$ but undetected at
24 \um\ (Figure \ref{seds}). More importantly, Figure \ref{seds} (as
well as inspection of the individual SEDs) shows that for most of our
objects the observed short-wavelength excess cannot explain their 24
\um\ emission, where the SED takes a significant turn upwards.
  
The low IR luminosity of these SMC stars in comparison to their
photospheric stellar luminosity is characteristic of systems hosting
debris disks \citep[e.g.,][]{rieke05,gorlova06,chen06}. These disks
are mostly devoid of gas, and form and replenish through the grinding
of planetesimals.  The small 24 \um\ excess in our SMC early-B stars
relative to their luminosity ($10^{-4}<L_{IR}/L_*<10^{-2}$) is very
similar to the typical excess observed in debris disks
($L_{IR}/L_*\lesssim10^{-3}$), where only a small fraction of the
radiation from the central source is captured by the dust and
reradiated in the IR \citep[e.g.,][]{uzpen05}. This is, of course, an
exciting possibility since it would provide indirect evidence for the
existence of planets in a galaxy other than our own.  There is a
well-known deficit of giant planets (found by radial velocity
techniques) around subsolar-metallicity stars that suggests such
planets would be rare in a low-metallicity galaxy such as the SMC
\citep{fischer05}. Recent observations, however, find that this
deficit of giant planets does not translate into a deficit of debris
disks \citep{greaves06,bryden06}.  Debris disks around massive early-B
stars would be populated by millimeter-size particles, as radiation
pressure should rapidly clear out smaller particles (unless gas drag
is significant). By comparison, recent spectroscopic observations
suggest that debris disks around less massive A and F stars are
populated by particles with sizes larger than 10 \um\
\citep{jura04,chen06}.

The final possibility we consider for the origin of the 24~\um\
emission is cirrus hotspots, sometimes noted as the Pleiades effect
\citep[e.g.,][]{sloan04}.  Cirrus hotspots occur where radiation from
a massive star dominates over the ISRF, heating a nearby patch of
interstellar dust above the typical temperature of cirrus emission
\citep{vanburen89}. Following \citet{vanburen88}, we estimate that the
distance over which an early-B star can heat small ($\sim0.1$~\um\
size) ISM dust grains to temperatures T$_{dust}\sim150$~K is
$r\sim0.01-0.3$~pc, or $r\lesssim1\arcsec$ at the distance of the SMC.
This would appear as a unresolved source in our data for everything
except the brightest B0-O9 stars, and thus be very difficult to
disentangle from a remnant accretion disk or a debris disk without
additional observations at higher resolution.

\section{Summary and Conclusions}
\label{conclusion}

We have used the {\em Spitzer Space Telescope} to image the
star-forming body of the SMC in the mid and far IR with unprecedented
sensitivity and resolution.

IRAC images of the SMC show a wealth of diffuse emission across the
SMC seen for the first time. Particularly striking is the diffuse
emission at 8 $\mu$m, often associated with star-forming regions.  We
have used these images to take a first look at the \fei/\ftf\ ratio
(an estimator of PAH abundance) across the SW Bar region of the SMC.
We find that this ratio has large spatial variations, and takes a wide
range of values even in a small region.  Extended emission in the
southernmost part of the SW Bar has $\fei/\ftf>0.7$, typical of
solar-metallicity galaxies, while near the sites of massive star
formation and bright 24 \um\ emission (suggesting large radiation
fields) the ratio drops to the $\fei/\ftf<0.1$ observed in metal-poor
galaxies \citep{engelbracht05}. Because existing surveys of abundances
in \hii\ regions find no evidence for a gradient (or even a large
scatter) for metallicity in the SMC, we conclude that the wide range
of values taken by the \fei/\ftf\ ratio in this galaxy, and the wide
range of PAH abundance suggested by it, is primarily driven by the
local ISRF and not metallicity. We find that the relationship between
the \fei/\ftf\ ratio and the FIR dust color represented by the
\ftf/\fst\ ratio can be described by $\log[\fei/\ftf]\approx
-0.7\log[\ftf/\fst]-1.4$, showing that regions of blue FIR color are
deficient in 8 \um\ PAH emission relative to their 24 \um\ intensity.

We combine these observations with optical and near-IR photometry in
the literature to produce a catalog of point sources that spans from
optical wavelengths to 70 $\mu$m. We use this catalog to investigate
the nature of the sources detected in the {\em Spitzer} wavebands.  We
find that the most prominent sources detected at the long wavelengths
fall into four categories: 1) sources with very faint optical
counterparts and very red ($[5.8]-[8.0]>1.2$) colors.  This
population corresponds to YSOs, with some contamination from
background galaxies and foreground dwarfs that we estimate using the
SWIRE ELAIS-N1 catalog.  2) Bright MIR sources ($-12<M_{5.8}<-7.5$) with
midly red colors ($0.16\lesssim[5.8]-[8.0]<0.6$), many of which have
bright optical counterparts. These are identified as carbon stars.  3)
Bright mid infrared sources with neutral colors and bright optical
counterparts, corresponding to oxygen-rich AGB and red supergiant
stars. And 4) a class of dusty early B stars (B3 to O9, based on the
optical photometry) that are not appreciably reddened optically, but
that have large excesses at 24 $\mu$m most of which cannot be
explained by free-free emission. We use our multiwavelength data to
compile a catalog of 282 bright YSOs in the SMC, where we choose the
color-magnitude cuts to allow for only 7\% contamination by background
or foreground sources (Table 4).

Finally, we discuss the population of early B stars detected at 24
\um, which constitutes a few percent ($\sim3-5\%$) of the B stars in a
narrow region of the optical color-magnitude space, and shares many of
the observational signatures of debris disks or cirrus hotspots in the
Milky Way. Based on their small near IR excess and the fact that most
of these objects are not found in existing objective prism catalogs,
we conclude that the bulk of these systems are not emission line
stars. They may constitute a population of young massive stellar
systems caught in the act of photoevaporating their accretion disks,
but the timescale derived by models for such process suggests that
this is unlikely. Dust associated with debris disks or cirrus hotspots
(i.e., the Pleiades effect) appear to be the most likely explanation
for the origin of their 24 \um\ emission.

The S$^3$MC point source catalog constitutes a deep and uniform
photometric database for a large sample of stars located at the same
distance, with similar metallicities, and low levels of foreground and
background confusion. We expect that one of the contributions of this
work will be to supply the basis on which to test different schemes
for identifying mid- and far-infrared sources. The S$^3$MC
observations provide a unique resource for the study of stellar
evolution, dust production, dust properties, and the interaction
between the ISM and star formation in an environment that, because of
its proximity and low metallicity, has closer resemblance to the
conditions prevalent in primitive galaxies than any other galaxy that
can be studied in similar detail.

\acknowledgements This work is based on observations made with the
\emph{Spitzer Space Telescope}, which is operated by the Jet
Propulsion Laboratory, California Institute of Technology under a
contract with NASA. This research was supported in part by NSF grant
AST-0228963. Partial support for this work was also provided by NASA
through an award issued by JPL/Caltech (NASA-JPL Spitzer grant 1264151
awarded to Cycle 1 project 3316).  JDS acknowledges support from a
Millikan Fellowship provided by Caltech. MR is supported by the
Chilean {\sl Center for Astrophysics} FONDAP No. 15010003.  This
research has made use of the VizieR catalog access tool and the SIMBAD
database, operated at CDS, Strasbourg, France. This research also took
extensive advantage of NASA's Astrophysics Data System Bibliographic
Services.  This publication makes use of data products from the Two
Micron All Sky Survey, which is a joint project of the University of
Massachusetts and the Infrared Processing and Analysis
Center/California Institute of Technology, funded by the National
Aeronautics and Space Administration and the National Science
Foundation. The authors acknowledge the thorough and enthusiastic
review by the referee, Greg Sloan, whose comments much improved the
paper.  We wish to especially thank R. Gruendl at the University of
Illinois at Urbana-Champaign and D. Makovoz at the SSC for their great
help in dealing with data reduction issues, and E. Chiang,
D. Hollenbach, C. McKee, M. Cohen, P. Kalas, and J. Graham at the
University of California, Berkeley for helpful and interesting
discussions. A. Bolatto wants to especially acknowledge the patience
and support from his wife, Liliana, who gave birth to Sof\'{\i}a
Eliana in the midst of this research.

\clearpage
\begin{figure}
\includegraphics[angle=90,height=9.5in]{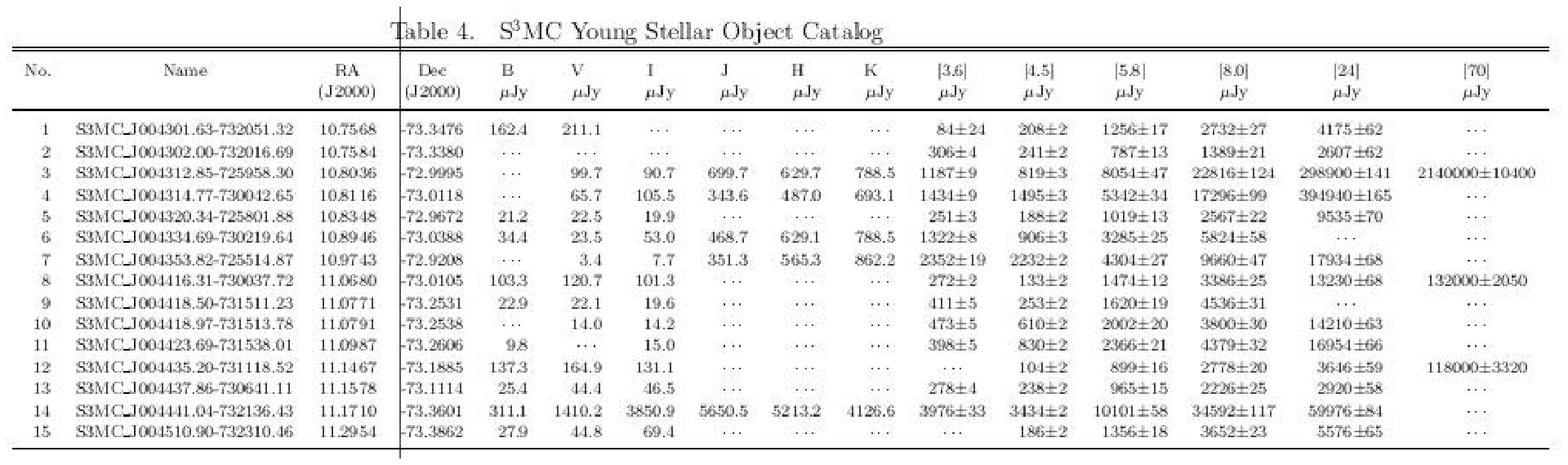}
\end{figure}

\end{document}